\documentclass[aps, prl, twocolumn, superscriptaddress,longbibliography,nobibnotes]{revtex4-1}

\usepackage{graphicx}
\usepackage[hidelinks]{hyperref} 
\usepackage{bm}
\usepackage{amsmath}
\usepackage{bbold}

\graphicspath{{./}}

\begin{document}


\title{Spinon continuum in the Heisenberg quantum chain compound Sr$_2$V$_3$O$_9$} 
\thanks{This manuscript has been authored by UT-Battelle, LLC under Contract No. DE-AC05-00OR22725 with the U.S. Department of Energy.  The United States Government retains and the publisher, by accepting the article for publication, acknowledges that the United States Government retains a non-exclusive, paid-up, irrevocable, world-wide license to publish or reproduce the published form of this manuscript, or allow others to do so, for United States Government purposes.  The Department of Energy will provide public access to these results of federally sponsored research in accordance with the DOE Public Access Plan (http://energy.gov/downloads/doe-public-access-plan).}

\renewcommand*{\thefootnote}{\arabic{footnote}}

\author{Shang Gao}
\email[]{sgao.physics@gmail.com}
\affiliation{Neutron Scattering Division, Oak Ridge National Laboratory, Oak Ridge, TN 37831, USA}
\affiliation{Materials Science \& Technology Division, Oak Ridge National Laboratory, Oak Ridge, TN 37831, USA}
\affiliation{Department of Physics, University of Science and Technology of China, Hefei, Anhui 230026, People's Republic of China}

\author{Ling-Fang Lin}
\affiliation{Department of Physics and Astronomy, University of Tennessee, Knoxville, TN 37996, USA}

\author{Pontus Laurell}
\affiliation{Department of Physics and Astronomy, University of Tennessee, Knoxville, TN 37996, USA}

\author{Qiang Chen}
\affiliation{Department of Physics and Astronomy, University of Tennessee, Knoxville, TN 37996, USA}

\author{Qing Huang}
\affiliation{Department of Physics and Astronomy, University of Tennessee, Knoxville, TN 37996, USA}

\author{Clarina dela Cruz}
\affiliation{Neutron Scattering Division, Oak Ridge National Laboratory, Oak Ridge, TN 37831, USA}

\author{Krishnamurthy V. Vemuru}
\affiliation{Neutron Scattering Division, Oak Ridge National Laboratory, Oak Ridge, TN 37831, USA}
\affiliation{BrainChip Inc., Laguna Hills, CA 92653, USA}

\author{Mark D. Lumsden}
\affiliation{Neutron Scattering Division, Oak Ridge National Laboratory, Oak Ridge, TN 37831, USA}

\author{Stephen E. Nagler}
\affiliation{Neutron Scattering Division, Oak Ridge National Laboratory, Oak Ridge, TN 37831, USA}
\affiliation{Department of Physics and Astronomy, University of Tennessee, Knoxville, TN 37996, USA}

\author{Gonzalo Alvarez}
\affiliation{Computational Sciences \& Engineering Division, Oak Ridge National Laboratory, Oak Ridge, TN 37831, USA}

\author{Elbio Dagotto}
\affiliation{Materials Science \& Technology Division, Oak Ridge National Laboratory, Oak Ridge, TN 37831, USA}
\affiliation{Department of Physics and Astronomy, University of Tennessee, Knoxville, TN 37996, USA}

\author{Haidong Zhou}
\affiliation{Department of Physics and Astronomy, University of Tennessee, Knoxville, TN 37996, USA}

\author{Andrew D. Christianson}
\affiliation{Materials Science \& Technology Division, Oak Ridge National Laboratory, Oak Ridge, TN 37831, USA}

\author{Matthew B. Stone}
\email[]{stonemb@ornl.gov}
\affiliation{Neutron Scattering Division, Oak Ridge National Laboratory, Oak Ridge, TN 37831, USA}

\date{\today}


\begin{abstract}
Magnetic excitations in the spin chain candidate Sr$_2$V$_3$O$_9$ have been investigated by inelastic neutron scattering on a single crystal sample. A spinon continuum with a bandwidth of $\sim22$~meV is observed along the chain formed by alternating magnetic V$^{4+}$ and nonmagnetic V$^{5+}$ ions. Incipient magnetic Bragg peaks due to weak ferromagnetic interchain couplings emerge when approaching the magnetic transition at $T_N\sim 5.3$~K while the excitations remain gapless within the instrumental resolution. Comparisons to the Bethe ansatz, density matrix renormalization group (DMRG) calculations, and effective field theories confirm Sr$_2$V$_3$O$_9$ as a host of weakly coupled $S = 1/2$ chains dominated by antiferromagnetic intrachain interactions of $\sim7.1$(1)~meV.
\end{abstract}

\maketitle

\section{I. Introduction}

Spin chains are one of the simplest models that illustrate many fundamental concepts in quantum magnets~\cite{giamarchi_quantum_2003}. The reduced number of neighboring sites greatly enhances quantum fluctuations and promotes exotic phenomena like fractional spinons~\cite{bethe_zur_1931,faddeev_what_1981} and valence bonds~\cite{affleck_rigorous_1987}. Compared to higher dimensional systems, an advantage of the chain models is that they can be solved with high accuracy~\cite{sutherland_beautiful_2004}. Starting from the Bethe ansatz for the $S = 1/2$ Heisenberg chains~\cite{bethe_zur_1931}, analytical or numerical solutions for spin chains have been obtained for various types of chains that incorporate perturbations like Ising anisotropy, interchain couplings, and magnetic fields, thus allowing a thorough understanding of a plethora of novel phenomena including Zeeman ladders~\cite{shiba_quantization_1980, coldea_quantum_2010, grenier_longitudinal_2015, mena_thermal_2020, lane_nonlinear_2020}, psinon excitations~\cite{karbach_line_2000, karbach_quasiparticles_2002}, and Bethe strings~\cite{wang_experimental_2018, bera_dispersions_2020}.

\begin{figure}[t!]
    \includegraphics[width=0.45\textwidth]{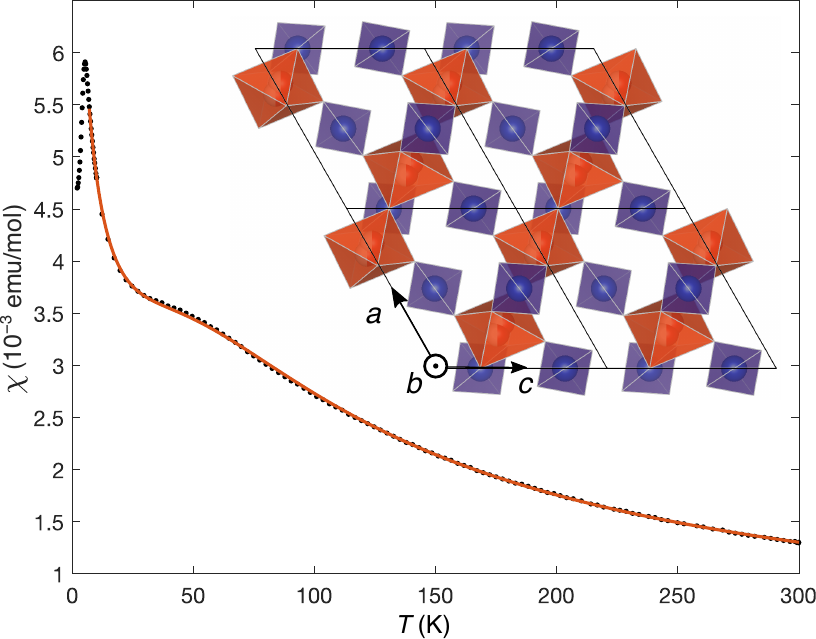}
    \caption{Magnetic susceptibility of Sr$_2$V$_3$O$_9$ measured in a 5~kOe field. An antiferromagnetic transition is observed at $T_N = 5.3$~K as a sharp peak. Red solid line is the fit to the $\chi(T)$ in the temperature range of [10, 300]~K, as described in the text. Inset is the Sr$_2$V$_3$O$_9$ crystallographic structure viewed along the $b$ axis. The V$^{4+}$O$_6$ octahedra and V$^{5+}$O$_4$ tetrahedra are shown in red and blue, respectively. Positions of the Sr$^{2+}$ ions are not shown for clarity.
    \label{fig:intro}}
\end{figure}

The strontium vanadate Sr$_2$V$_3$O$_9$ has been proposed as a host of the $S = 1/2$ Heisenberg antiferromagnetic chain~(HAFMC)~\cite{kaul_sr_2003, ivanshin_esr_2003}. Sr$_2$V$_3$O$_9$ belongs to the monoclinic $C2/c$ space group, with lattice constants determined as $a = 7.55$, $b= 16.28$, $c = 6.95$~\AA, and $\beta = 119.78^{\circ}$~\cite{mentre_structural_1998}. In this compound, the V-O layers in the $ac$ planes are separated at a large distance of $\sim8.14$~\AA\ by the Sr layers along the $b$ axis. As shown in the inset of Fig.~\ref{fig:intro}, within the V-O layers, the V$^{4+}$O$_6$ octahedra containing the magnetic V$^{4+}$ ions ($S = 1/2$) share corners along the $\bm{a}+\bm{c}$ direction. Along the $\bm{a}-\bm{c}$ direction, the V$^{4+}$O$_6$ octahedra are linked across the nonmagnetic V$^{5+}$O$_4$ tetrahedra. Surprisingly, thermal transport measurements on a crystal sample indicate the spin chains are along the $\bm{a}-\bm{c}$ direction~\cite{kawamata_thermal_2014}, suggesting stronger spin couplings across the nonmagnetic V$^{5+}$O$_4$ tetrahedra. Although such a scenario was supported by the density functional theory (DFT) calculations~\cite{rodriguez_first_2010}, direct spectroscopic evidence for chain physics in Sr$_2$V$_3$O$_9$ is still missing.

Here we utilize neutron scattering to study the spin dynamics in Sr$_2$V$_3$O$_9$. A gapless spinon continuum, which is a characteristic feature of the $S=1/2$ Heisenberg chain, is observed at temperatures down to $\sim5$~K. The chain direction is determined to be along the $\bm{a}-\bm{c}$  direction, thus verifying the scenario deduced from the thermal transport experiments~\cite{kawamata_thermal_2014}. By comparing the inelastic neutron scattering (INS) spectra with the Bethe ansatz, density matrix renormalization group (DMRG) calculations, and field theories, we conclude Sr$_2$V$_3$O$_9$ is a host of weakly coupled $S = 1/2$ HAFMCs.

\section{II. Methods}

Sr$_2$V$_3$O$_9$ crystals were prepared using a floating zone image furnace following reported procedures~\cite{uesaka_thermal_2010}. In order to synthesize phase pure Sr$_2$V$_3$O$_9$, polycrystalline Sr$_2$V$_2$O$_7$ was first prepared using a stoichiometric SrCO$_3$ and V$_2$O$_5$ powder mixture fired at 700$^{\circ}$C for 72 hours in air. The obtained Sr$_2$V$_2$O$_7$ powder was then mixed with VO$_2$ powder in a molar ratio of 1:1. The mixture was pressed into a rod of $\sim7$~mm in diameter, $\sim10$~cm in length, and then annealed at 540$^{\circ}$C in argon for 24 hours. The following floating zone growth was performed using a NEC two-mirror image furnace. As is reported in Ref.~\cite{uesaka_thermal_2010}, the twice-scanning technique is utilized for this growth. The first scan was a fast scan with a speed of 35~mm/h under flowing Ar of 2.5~atm. The 2nd growth scan, was done using a speed of 1~mm/h in the same gas flow. Several large segments of single crystal were obtained. These crystals were then oriented by backscattering X-ray Laue diffraction in preparation for the neutron scattering measurements. DC magnetic susceptibility measurements were performed at temperatures of 2-300~K using a Quantum Design superconducting quantum interference device - Vibrating Sample Magnetometer (SQUID-VSM). The sample is cooled in zero-field (ZFC) and measured in an external field of 0.5 T for increasing temperatures. 

Inelastic neutron scattering (INS) experiments on Sr$_2$V$_3$O$_9$ were performed on the fine-resolution Fermi chopper spectrometer SEQUOIA at the Spallation Neutron Source (SNS) of the Oak Ridge National Laboratory (ORNL). A single crystal with mass of $\sim$200 mg was aligned with the $b$ axis vertical. A closed cycle refrigerator (CCR) was employed to reach temperatures, $T$, down to 5 K. Incident neutron energies were $E_i$ = 35, 10, and 4~meV. For the $E_i = 35$~meV measurements, a Fermi chopper frequency of 240~Hz was used with the high flux chopper. Data were acquired by rotating the sample in 1$^\circ$ steps about its vertical axis, covering a total range of 165$^{\circ}$ at $T = 5$, 20, and 50~K. For the $E_i = 10$ and 4~meV measurements, a Fermi chopper frequency of 120~Hz was used with the high resolution chopper. Data for the $E_i = 10$~meV (4~meV) measurements at 4~K were acquired by rotating the sample in 1$^\circ$ (0.4$^\circ$) steps, covering a total range of 200$^\circ$ (39.2$^\circ$). Measurements of an empty sample holder were subtracted as the background. Data reductions and projections were performed using the MANTID software~\cite{arnold_mantid_2014}.

For the theoretical calculations, the canonical one-dimensional isotropic $S = 1/2$ HAFMC model described by the Hamiltonian $\mathcal{H} = J\sum_{\langle \textrm{NN}\rangle}\bf{S}_i \cdot \bf{S}_j$ is adopted, where the summation is over the nearest neighbors (NN). The $T=0$ dynamical spin structure factor was calculated in the algebraic Bethe ansatz approach using the ABACUS algorithm~\cite{caux_correlation_2009}. The calculation was performed on a system of $L=500$ sites with periodic boundary conditions, using an energy step of $\Delta \omega = 0.002J$. A Gaussian energy broadening of $0.02J\approx 0.142$~meV was applied. A sum rule saturation of $99\%$ was reached, which can be compared with the approximately $98\%$ saturation expected from the two- and four-spinon contributions to the total intensity in the thermodynamic limit~\cite{caux_four_2006}.

Theoretical spectra were also calculated using the density matrix renormalization group (DMRG) technique~\cite{white_density_1992,white_density_1993} as implemented in the DMRG++ code~\cite{alvarez_density_2009}. The calculations were carried out using the Krylov-space correction vector approach~\cite{kuhner_dynamical_1999, nocera_spectral_2016} with open boundary conditions (OBC). Targeting a truncation error below $10^{-10}$, a minimum of 100 and up to 1000 states were kept during our DMRG calculations.  The half width at half maximum of the Lorentzian energy broadening was set as $0.1J$.  For the $T = 0$ DMRG calculations, we used a chain with $N = 100$ sites, while for the $T >0$ calculations we adopted a system of 50 physical and 50 ancilla sites by using the ancilla (or purification) method~\cite{feiguin_finite_2005,feiguin_spectral_2010, nocera_symmetry_2016}. Examples of input files and more details can be found in the Supplemental Materials~\cite{supp}.

\section{III. Results and Discussions}

As a reference for the temperature evolution of the spin correlations, Fig.~\ref{fig:intro} presents the magnetic susceptibility $\chi(T)$ measured on pulverized single crystals of Sr$_2$V$_3$O$_9$. A broad hump around $T = 50$~K signals strong short-range spin correlations. Following Ref.~\cite{kaul_sr_2003}, we fit $\chi(T)$ to $\chi_{\rm{1D}} + \chi_{\rm{LT}} + \chi_{\rm{vv}}$, where $\chi_{\rm{1D}}$ is the polynomial approximation of the contribution from a $S=1/2$ Heisenberg chain~\cite{Feyerherm_magnetic_2000}, $\chi_{LT}$ is a Curie-Weiss term to account for the upturn at low temperatures, and $\chi_{\rm{vv}}$ is a temperature-independent Van Vleck contribution. The fitted intrachain coupling strength is $J=6.95(5)$~meV, which is close to the previously  reported value~\cite{kaul_sr_2003}. An antiferromagnetic transition is observed at $T_N\sim5.3$~K,  indicating the existence of weak interchain couplings $J_\perp$. A tiny jump in $\chi(T)$ above $T_N$, which is described by the $\chi_{\rm{LT}}$ term, has been ascribed to the antisymmetric Dyaloshinskii-Moriya (DM) interactions~\cite{kaul_sr_2003,ivanshin_esr_2003}, although such a scenario cannot be directly verified in our zero-field experiments~\cite{dender_direct_1997, oshikawa_field_1997}.

\begin{figure}[t!]
    \includegraphics[width=0.48\textwidth]{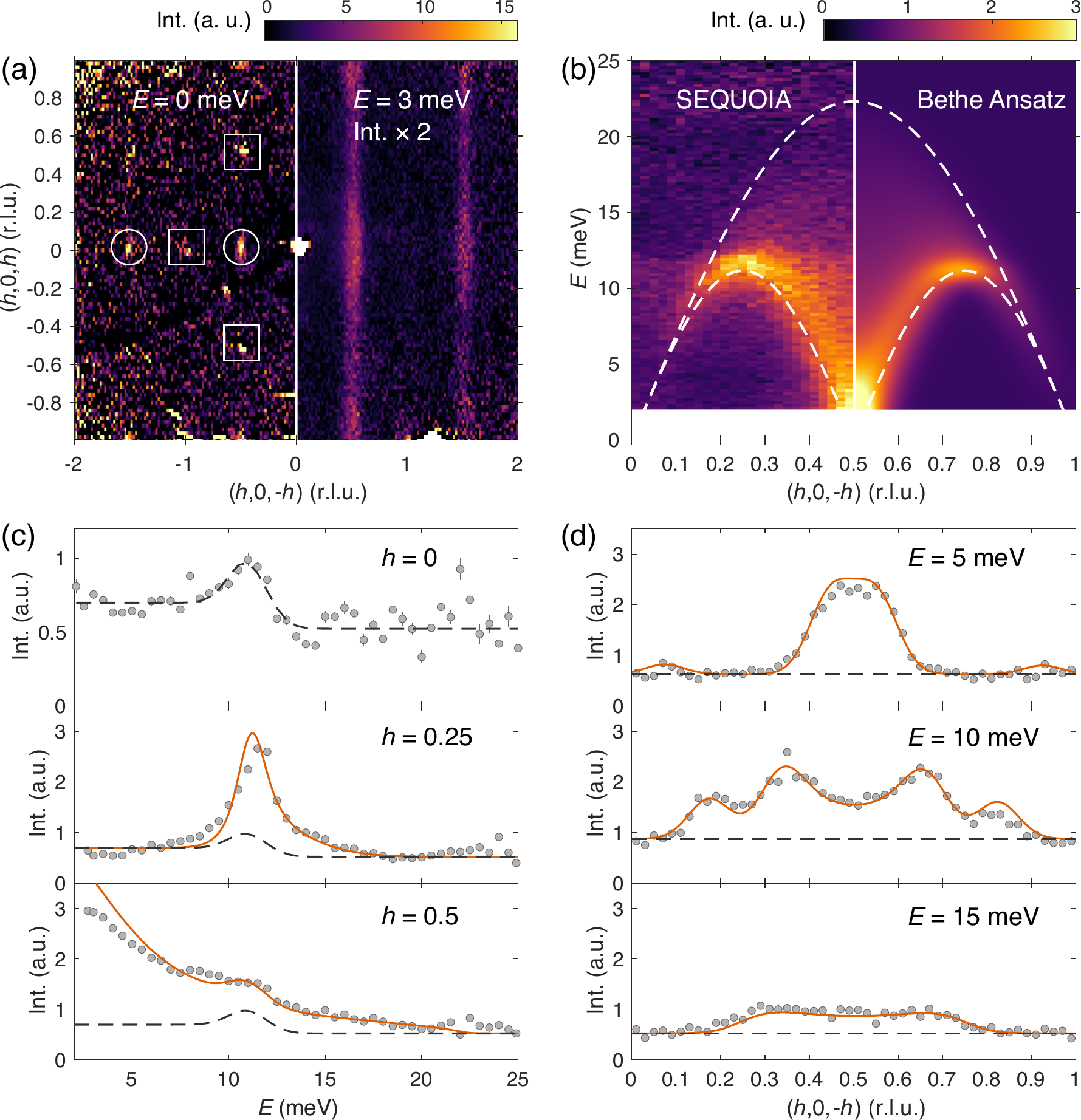}
    \caption{(a) Constant energy slices of the Sr$_2$V$_3$O$_9$ INS spectra $S(Q,\omega)$ in the $(h,0,l)$ plane at $E=0$ and 3~meV integrated in the energy range of $[-2, 2]$ and $[2,4]$~meV, respectively. Data is plotted on an orthogonal coordinate system for clarity. The incident neutron energy is $E_i=35$~meV with the measuring temperature $T=5$~K. Intensity in the slice at $E = 3$~meV is multiplied by a factor of 2 for better visibility. In the $E=0$~meV slice, circles and squares indicate the magnetic and nuclear Bragg peaks, respectively. (b) Comparison between the experimental and theoretical $S(Q,\omega)$ along the $(h,0,\bar{h})$ direction. The experimental data is integrated in the range of $\delta h=[-1.2, 1.2]$~r.l.u. along $(h,0,h)$. For the calculated cross section, a constant intensity is added to account for any measurement background. Dashed lines are the lower and upper boundaries of the 2-spinon continuum for $J = 7.1$~meV. (c) Scattering intensity as a function of $E$ at $h$ = 0, 0.25, and 0.5 along $(h,0,\bar{h})$. Black dashed line is a fit to the background scattering at $h=0$ using a Gaussian function plus a step function. Red solid line is a fit to the spinon continuum plus the background. (d) Scattering intensity as a function of $(h,0,\bar{h})$ at $E=5$, 10, and 15~meV. Black dashed line indicates the constant background extracted from the scan at $h=0$ in panel (c). Red solid line is a fit to the spinon continuum plus the background.
    \label{fig:SEQ}}
\end{figure}

The existence of a magnetic ordered state below $T_N$ is directly confirmed by the neutron scattering data measured at $T = 5$~K. For the elastic map shown in the left panel of Fig.~\ref{fig:SEQ}(a), the data collected at $T = 50$~K is subtracted to expose the weak magnetic reflections. Unless otherwise stated, all presented neutron scattering data are integrated along the (0,$k$,0) direction in the range of $k = [-4,4]$ reciprocal lattice units (r.l.u.) to improve counting statistics. Therefore, the strongest magnetic reflection in Fig.~\ref{fig:SEQ}(a), $(1/2,k,-1/2)$, can be indexed as $(1/2, 1, -1/2)$, revealing the magnetic propagation vector to be $\bm{q} = (1/2, 0, 1/2)$. As this $\bm{q}$ vector indicates parallel spin alignment along the $\bm{a}+\bm{c}$ direction, we can conclude that the weak interchain coupling should be ferromagnetic if assuming the strongest interchain couplings arise from the corner-sharing V$^{4+}$O$_6$ octahedra along the $\bm{a}+\bm{c}$ direction.

At an energy transfer of $E = 3$~meV, the constant energy map shown in the right panel of Fig.~\ref{fig:SEQ}(a) exhibits narrow streaks along the $(h,0,h)$ direction. Such a highly anisotropic scattering pattern is direct evidence for the emergence of spin chains in Sr$_2$V$_3$O$_9$, with chains running along the $\bm{a}-\bm{c}$ direction. Weak modulation along the streaks can be ascribed to the perturbations from the interchain couplings. In the Supplemental Materials~\cite{supp}, we present slices along the $(0,k,0)$ the $(h,0,h)$ directions, which reveal very weakly dispersive excitations due to marginal interchain couplings.

\begin{figure}[t!]
    \includegraphics[width=0.45\textwidth]{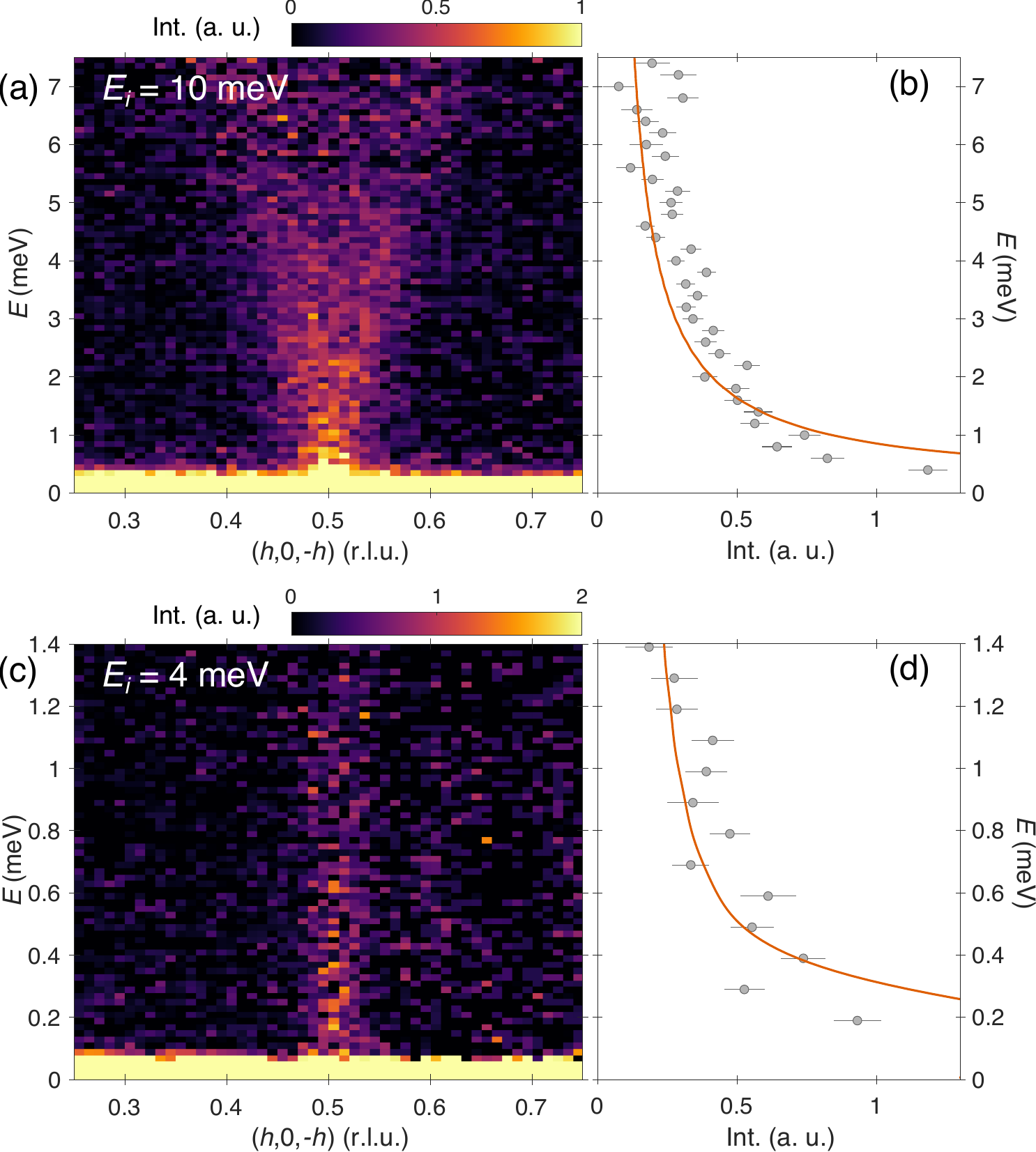}
    \caption{(a) Low-energy section of the INS spectra $S(Q,\omega)$ measured with an incident neutron energy of $E_i = 10$~meV. Data along directions perpendicular to the chain have been integrated for better statistics. (b) Scattering intensity as a function of $E$ integrated in the range of $\delta h = [0.49, 0.51]$ along $(h,0,-h)$. Red line is the theoretical scattering cross section for a HAFMC plus a constant background. (c),(d) Similar as panels (a),(b) with a lower $E_i = 4$~meV.
    \label{fig:gap}}
\end{figure}

After integrating the INS data along the $(h,0,h)$ direction within a range of $\delta h = [-1.2, 1.2]$ r.l.u., the excitation spectra along the $(h,0,\bar{h})$ direction are obtained. As shown in the left panel of Fig.~\ref{fig:SEQ}(b), the spectra exhibit a continuum of excitations up to $\sim 22$~meV, which is a typical feature of fractional spinon excitations of HAFMCs~\cite{muller_quantum_1981, tennant_measurement_1995,lake_quantum_2005, mourigal_fractional_2013, wu_tomonaga_2019}. The lower and upper boundaries of the 2-spinon continuum can be described by $\omega_L(q) = (\pi/2)J|\sin q|$ and $\omega_U(q) = \pi J|\sin(q/2)|$, respectively, where $J$ is the strength of the intrachain couplings~\cite{muller_quantum_1981}. To describe the shape of the spinon continuum in Sr$_2$V$_3$O$_9$, $J = 7.1(1)$~meV is determined by a $\chi^2$ fit of the spinon continuum, and the corresponding boundaries are overplotted in Fig.~\ref{fig:SEQ}(b) as dashed lines. Weak scattering intensities are observed outside the continuum boundary, including a step like excitation below $\sim12$~meV and a broad flat band around $\sim23$~meV. Since these features exhibit no wavevector dependence~\cite{supp}, they may be ascribed to the background scattering due to possible oxygen deficiency and the consequent valence variance of the vanadium ions.

Various analytical and numerical methods have been developed to describe the dynamical structure factor of the HAFMCs. Here we first compare the INS spectra of Sr$_2$V$_3$O$_9$ to the cross section calculated by the Bethe ansatz. The 2-spinon continuum is known to account for $\sim71~\%$ of the total spectral weight~\cite{bougourzi_exact_1996, karbach_two_1997, caux_four_2006, mourigal_fractional_2013}, while the remaining spectral weight is mostly accounted for by the 4-spinon continuum~\cite{caux_four_2006, mourigal_fractional_2013, lake_multispinon_2013}. As a zero temperature calculation method, the comparison with the experimental data acquired at 5~K is justified since the overall bandwidth of the system, which sets the relevant energy scale of the 1D fluctuations, is at much higher energy scales than the measuring temperature.

For $J = 7.1$~meV, the calculated spectral function is shown on the right panel in Fig.~\ref{fig:SEQ}(b). The calculated data are convolved by a Gaussian function with a full-width-half-maximum of $\Delta Q = 0.1$~r.l.u. along the $Q$ axis and by the instrumental energy resolution along the $E$ axis~\cite{supp}. More detailed comparisons for scans at constant $Q$ and $E$ are presented in Fig.~\ref{fig:SEQ}(c) and (d), respectively. For the background scan at $h = 0$, the intensity is fitted by a Gaussian function plus a step function to account for the additional scattering at $\sim 12$~meV described earlier. This is then added to the other calculated spectra shown in Fig.~\ref{fig:SEQ}(c) and (d). The calculation reproduces the INS spectra, thus confirming the existence of HAFMCs in Sr$_2$V$_3$O$_9$.

\begin{figure}[t]
    \includegraphics[width=0.49\textwidth]{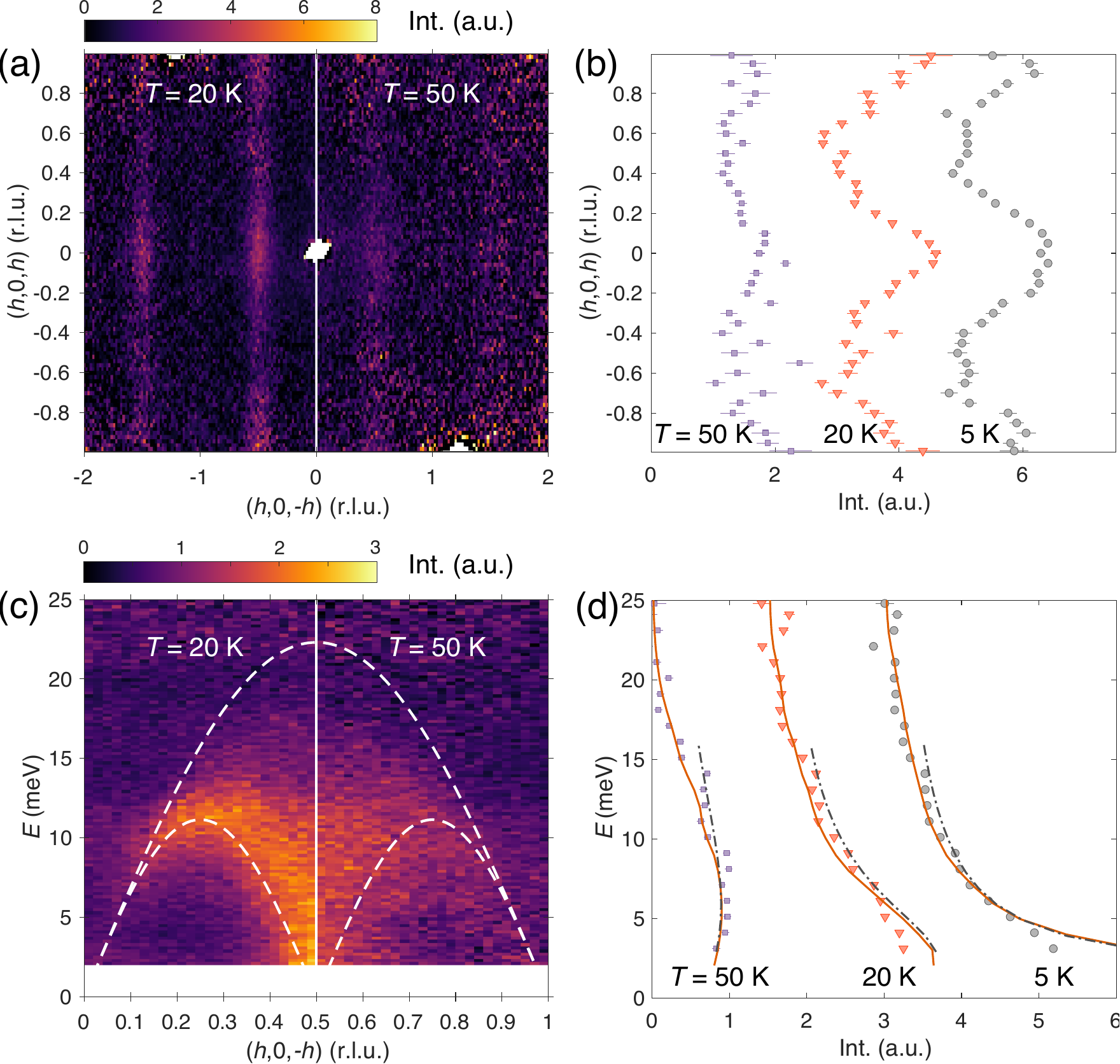}
    \caption{(a) Constant energy slices of the Sr$_2$V$_3$O$_9$ INS spectra $S(Q,\omega)$ in the $(h,0,l)$ plane measured at $T = 20$ (left) and 50~K (right). Data is integrated in the energy range of $[2,4]$~meV and is plotted on an orthogonal coordinate system for clarity. (b) Comparison of the scattering intensity along $(h,0,h)$ measured at $T = 5$ (gray circles), 20 (red triangles), and 50~K (purple squares). Data is integrated in the range of $\delta h=[0.45, 0.55]$ r.l.u.~along $(h,0,-h)$ and $E=[2,4]$~meV. Data at 20 (50)~K is shifted along the $x$ axis by 1.5 (3) for clarity.  (c) $S(Q,\omega)$ along $(h,0,-h)$ measured at $T=20$ (left) and 50~K (right). Data is integrated in the range of $\delta h=[-1.2, 1.2]$ r.l.u. along $(h,0,h)$ and $\delta k=[-4,4]$ r.l.u. along $(0,k,0)$. Dashed lines are the lower and upper boundaries of the 2-spinon continuum for J = 7.1 meV. (d) Comparison of the scattering intensity as a function of $E$ measured at $T = 5$ (gray circles), 20 (red triangles), and 50~K (purple squares). Data is integrated in the range of $\delta h = [0.475, 0.525]$ along $(h,0,-h)$. At each temperature, the spectra at $Q = 0$ measured at $T = 5$~K is subtracted as the background. Red solid (black dash-dotted) lines are theoretical scattering cross section calculated by DMRG (field theory) at $T$ = 50, 20, and 0~K (50, 20, and 5~K) assuming $J = 7.1$~meV. Data at 20 (50)~K is shifted along the $x$ axis by 1.5 (3) units for clarity.
    \label{fig:Tdep}}
\end{figure}

The obtained strength of the intrachain coupling of $J = 7.1(1)$~meV, together with the magnetic long-range order transition temperature $T_N$= 5.3~K, allows an estimate of the strength of the ferrromagnetic interchain coupling $J_\perp$. Following the mean field analysis~\cite{kaul_sr_2003, schulz_dynamics_1996}, $|J_\perp|$ is estimated to be $\sim0.16$~meV, which is $\sim2.3~\%$ of the intrachain coupling $J$. This agrees well with the extent of the dispersion measured orthogonal to the chain direction~\cite{supp}.

In order to resolve a possible gap in the spinon excitations, further INS experiments were performed with lower incident energies of $E_i = 10$ and 4~meV. Figure~\ref{fig:gap} summarizes the spectra after full integrations along directions perpendicular to the chain. As compared in Fig.~\ref{fig:gap}(b) and (d), the spectra at $(1/2,0,-1/2)$ follows the theoretical dynamical structure factor down to $\sim0.1$~meV. Therefore, it can be concluded that the spinon excitations in Sr$_2$V$_3$O$_9$ are gapless within the instrumental resolution of $\sim0.1$~meV.

The temperature evolution of the INS spectra is summarized in Fig.~\ref{fig:Tdep}. For $T = 20$ and 50~K, the constant-$E$ map at an energy transfer of $E = 3$~meV is compared in Fig.~\ref{fig:Tdep}(a). The main features are similar to the map at $T= 5$~K shown in Fig.~\ref{fig:SEQ}(a), but the scattering intensity is weaker at elevated temperatures. Figure~\ref{fig:Tdep}(b) compares the intensity along the $(h,0,h)$ direction at $T = 5$, 20 and 50~K. The intensity contrast along the streaks is reduced at elevated temperatures as thermal fluctuations overcome the interchain couplings.

After integration in the range of $\delta h=[-1.2, 1.2]$~r.l.u. along the $(h,0,h)$ direction, the spectral functions along $(h,0,-h)$ are compared in Fig.~\ref{fig:Tdep}(c) for $T = 20$ and 50~K. The dashed lines outline the 2-spinon continuum for $J = 7.1$~meV as in Fig.~\ref{fig:SEQ}(b). Besides the reduced scattering intensities, the excitations become softened at elevated temperatures, with a significant fraction of the scattering intensity lying below $\omega_L(q)$ at $T = 50$~K. 

According to theoretical calculations~\cite{starykh_dynamics_1997}, an intensity transfer from $(h=1/2, \omega \rightarrow 0)$ to $(h=0, \omega \rightarrow 0)$ is expected in the spectra function $S(q,\omega)$ at elevated temperatures due to thermal fluctuations. Although such an intensity transfer is not directly probed in our experiment, it may induce a peak at nonzero energies in the constant-$Q$ scan at $h=0.5$ as the zero energy intensity is greatly reduced. Figure~\ref{fig:Tdep}(d) compares the constant-$Q$ scans at $h=0.5$ for $T = 5$, 20, and 50~K. Theoretical spectral functions calculated by the DMRG method at the corresponding temperatures are plotted as red solid lines, which reproduce the spectral function over a large range of energy transfers. At $T =50$~K, the reduced intensities around $E=0$ is consistent with the theoretical prediction of the HAFMCs, thus confirming the chain physics in Sr$_2$V$_3$O$_9$.

The temperature evolution of the scattering intensity for $S = 1/2$ HAFMCs has also been investigated through effective field theories in the continuum limit~\cite{schulz_phase_1986, tennant_unbound_1993}. At relatively low energy transfers, the energy dependence of the cross section at $\bm{q} = (1/2, 0, -1/2)$ is expressed as
\begin{equation}
S(\omega) \propto (n_\omega +1)\textrm{Im}\left\{ {\frac{1}{T}\left[\rho\left(\frac{\omega}{4\pi k_BT}\right)\right]^2}\right\} \textrm{,}
\end{equation}
with the $\rho(x)$ function defined as
\begin{equation}
    \rho(x) = \frac{\Gamma(\frac{1}{4}-ix)}{\Gamma(\frac{3}{4}-ix)} \textrm{.}
\end{equation}
In this expression, $n_\omega$ is the Bose factor and $\Gamma$ is the complex gamma function. Using this expression, we calculate the cross section for $S = 1/2$ HAFMCs for energies up to 16 meV. The calculated results, with a fitted scale factor, are shown in Fig.~\ref{fig:Tdep}(d) as dash-dotted lines. In the calculated energy range, the field theoretical results capture the temperature evolution of both the experimental data and and the DMRG results, which further justifies the existence of HAFMCs in Sr$_2$V$_3$O$_9$.

\section{IV. Conclusions}

The existence of $S = 1/2$ HAFMCs in Sr$_2$V$_3$O$_9$ is spectroscopically confirmed through inelastic neutron scattering experiments and comparison with numerical simulations and mean field approximations. A spinon continuum is observed along the $(h,0,\bar{h})$ direction, verifying that the intrachain couplings are mediated by the nonmagnetic V$^{5+}$ ions. The spinon continuum, with a bandwidth of $\sim22$~meV, indicates the strength of the intrachain couplings to be $\sim7.1$(1)~meV. Despite the magnetic transition at $T_N\sim 5.3$~K, the excitations in Sr$_2$V$_3$O$_9$ remain gapless down to 5~K. Through comparisons to the Bethe ansatz, the density matrix renormalization group (DMRG) calculations, and the field theories, we conclude that Sr$_2$V$_3$O$_9$ is a host of weakly coupled $S = 1/2$ HAFMCs.

\begin{acknowledgments}
This work was supported by the U.S. Department of Energy, Office of Science, Basic Energy Sciences, Materials Sciences and Engineering Division. This research used resources at the Spallation Neutron Source (SNS) and the High Flux Isotope Reactor (HFIR), both are DOE Office of Science User Facilities operated by the Oak Ridge National Laboratory (ORNL). The work of S. N. and G. A. was supported by the U.S. Department of Energy, Office of Science, National Quantum Information Science Research Centers, Quantum Science Center. Q.C., Q.H, and H.Z. thank the support from National Science Foundation with Grant No. NSF-DMR-2003117.
\end{acknowledgments}

%

  \clearpage
  \newpage
  
  \renewcommand{\thefigure}{S\arabic{figure}}
  \renewcommand{\thetable}{S\arabic{table}}
  \renewcommand{\theequation}{S\arabic{equation}}

  \makeatletter
  \renewcommand*{\citenumfont}[1]{S#1}
  \renewcommand*{\bibnumfmt}[1]{[S#1]}
  \def\clearfmfn{\let\@FMN@list\@empty}  
  \makeatother
  \clearfmfn

  \setcounter{figure}{0} 
  \setcounter{table}{0}
  \setcounter{equation}{0} 
  
  \onecolumngrid
  \begin{center} {\bf \large Supplemental Materials for:\\Spinon continuum in the Heisenberg quantum chain compound Sr$_2$V$_3$O$_9$} \end{center}
  
  \vspace{0.5cm}

\renewcommand*{\thefootnote}{\arabic{footnote}}
\renewcommand{\thefigure}{S\arabic{figure}}
\renewcommand{\thetable}{S\arabic{table}}
\renewcommand*{\thefootnote}{\arabic{footnote}}

\section{Background scattering}
\begin{figure}[b!]
    \includegraphics[width=0.9\textwidth]{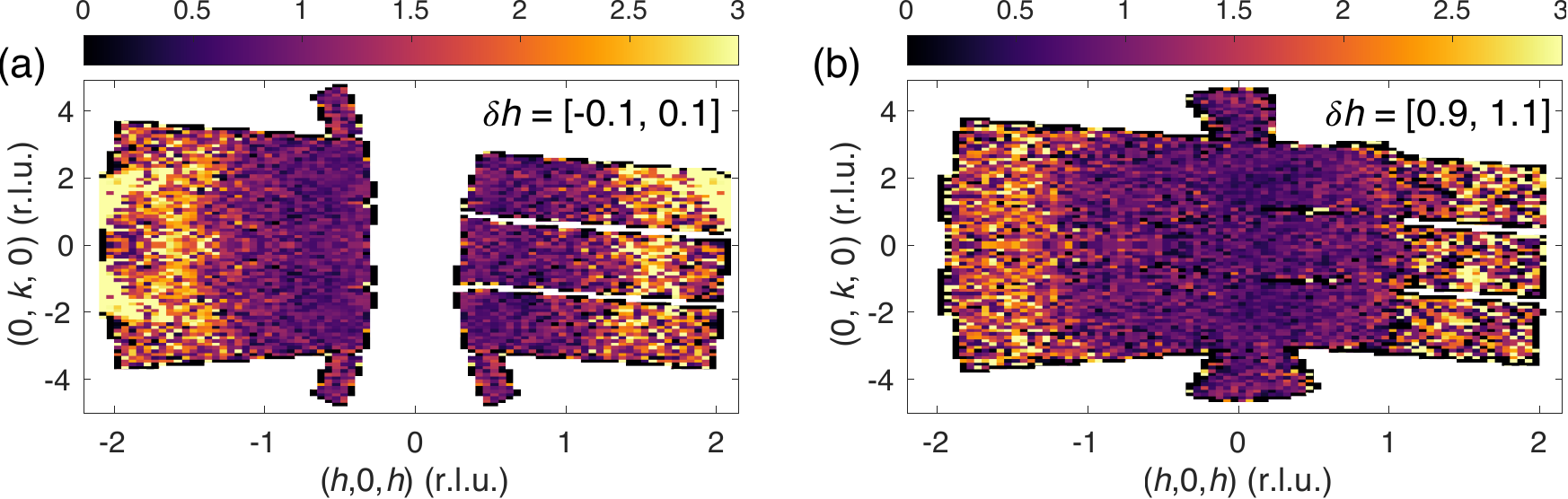}
    \caption{Constant energy slices at $E = 10.5$~meV integrated in the energy range of $\delta E = [8.5, 12.5]$~meV for data collected at $T = 5$~K with $E_i = 35$~meV. Data is integrated along the $(h,0,-h)$ direction in the range of $\delta h = [-0.1, 0.1]$ r.l.u. (a) and $\delta h = [0.9, 1.1]$ r.l.u. (b).
    \label{figs:bkgLE}}
\end{figure}

\begin{figure}[b!]
    \includegraphics[width=0.9\textwidth]{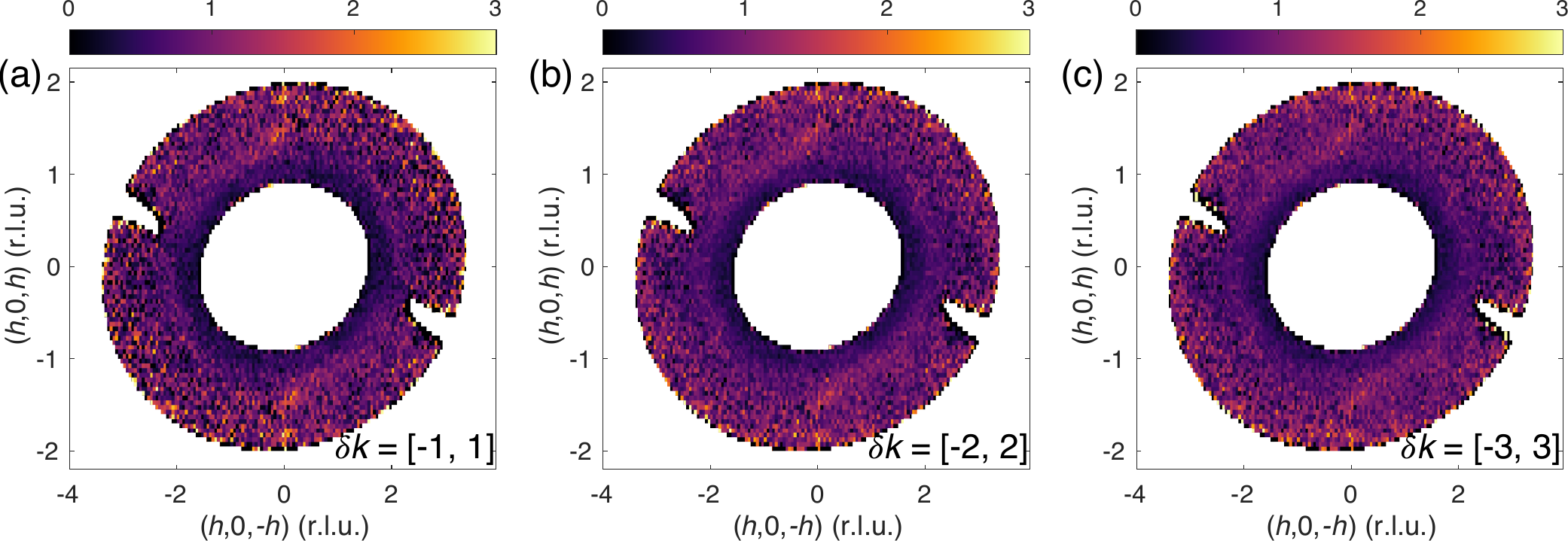}
    \caption{Constant energy slices at $E = 23.25$~meV integrated in the energy range of $\delta E = [22.0, 24.5]$~meV for data collected at $T = 5$~K with $E_i = 35$~meV. Data is integrated along the $(0,k,0)$ direction in the range of $\delta k = [-1, 1]$ r.l.u., and is plotted on an orthogonal coordinate system for clarity. (a) and $\delta k = [-2, 2]$ r.l.u. (b) and $\delta k = [-3, 3]$ r.l.u. (c).
    \label{figs:bkgHE}}
\end{figure}

To illustrate the wave vector independence of the background scattering around the energy transfer of $E \sim 11$ and 23~meV, Figs.~\ref{figs:bkgLE} and \ref{figs:bkgHE} present the constant energy slices within an energy transfer range of $\delta E = [8.5, 12.5]$~meV and $[22.0, 24.5]$~meV, respectively. For the slices in Fig.~\ref{figs:bkgLE}, data has been integrated in the range of $\delta h = [-0.1, 0.1]$ and [0.9, 1.1] r.l.u. along the $(h,0,-h)$ direction to stay away from the spinon continuum excitations. For the slices in Fig.~\ref{figs:bkgHE}, the energy transfer is higher than the upper boundary of the spinon continuum, therefore data is integrated in the range of $\delta k = [-1, 1]$,  $[-2, 2]$, and $[-3, 3]$ r.l.u. along the $(0,k,0)$ direction. No systematic wave vector dependence is observed in these background scattering slices.

\begin{figure}[t!]
    \includegraphics[width=0.9\textwidth]{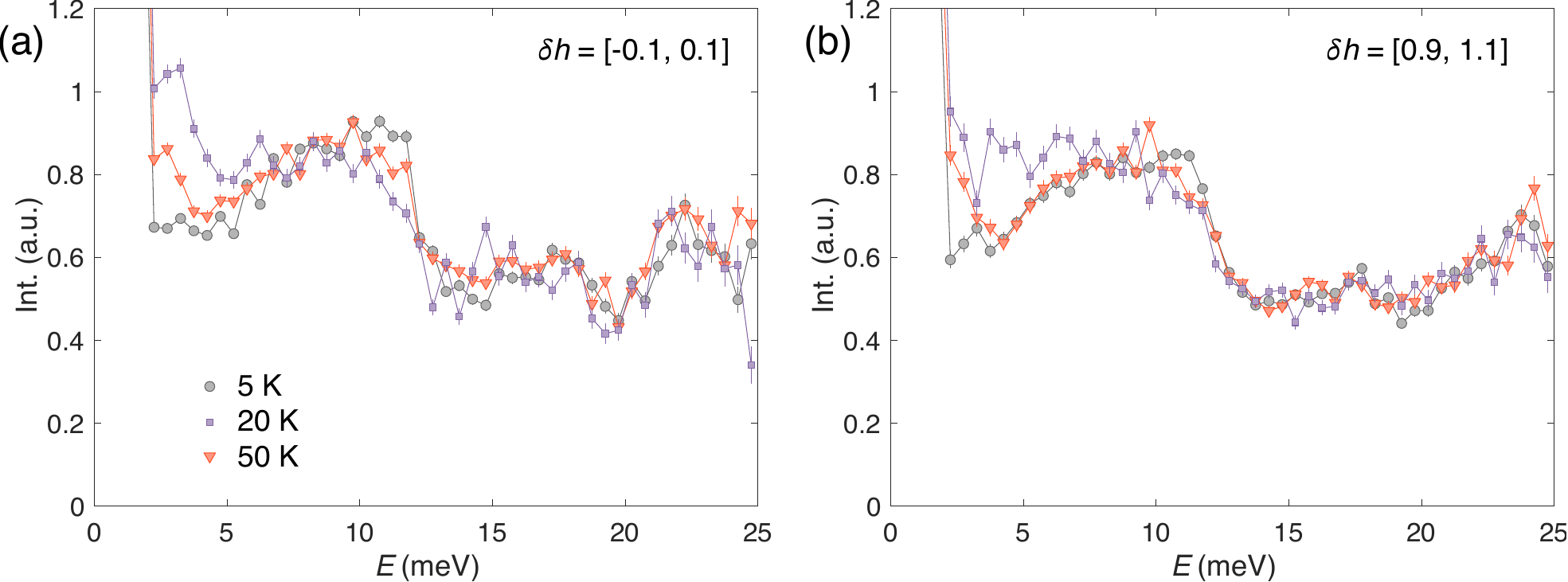}
    \caption{Scattering intensity as a function of $E$ integrated in the range of $\delta h = [-0.1, 0.1]$ r.l.u. (a) and $[0.9, 1.1]$ r.l.u. (b) along $(h,0,-h)$. Data is integrated in the range of $\delta k = [-4, 4]$ r.l.u. along $(0,k,0)$ and $\delta h = [-1.2, 1.2]$ r.l.u. along $(h,0,h)$.
    \label{figs:bkgT}}
\end{figure}

The temperature dependence of the background scattering is presented in Fig.~\ref{figs:bkgT}. Data is integrated in the range of $\delta h = [-0.1, 0.1]$ and $[0.9, 1.1]$ in Fig.~\ref{figs:bkgT} (a) and (b), respectively. At $E<5$~meV, intensity increases at elevated temperatures due to the intensity shift of the spinon continuum from $(h=1/2, \omega \rightarrow 0)$ to $(h=0, \omega \rightarrow 0)$ as discussed in the main text. The intensity from the background scattering at $E\sim11$ and 23~meV almost stays constant with temperature.

\begin{figure}[t!]
    \includegraphics[width=0.8\textwidth]{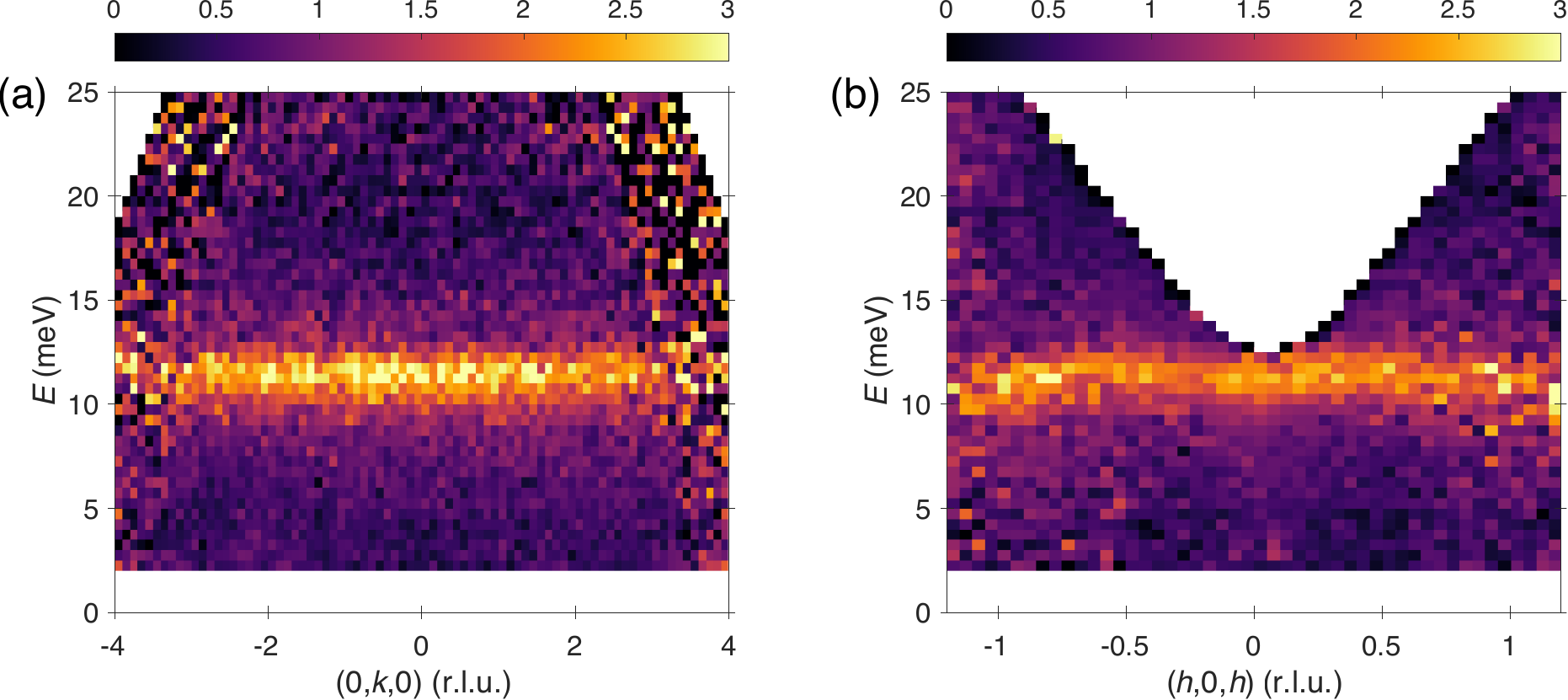}
    \caption{(a) INS spectra along the $(0,k,0)$ direction measured at $T = 5$~K. Data is integrated in the range of $\delta h = [0.2, 0.3]$ r.l.u. along $(h,0,-h)$ and $\delta h = [-1.2, 1.2]$ r.l.u. along $(h,0,h)$. (b) INS spectra along the $(h,0,h)$ direction measured at $T = 5$~K. Data is integrated in the range of $\delta h = [0.7, 0.8]$ r.l.u. along $(h,0,-h)$ and $\delta k = [-4, 4]$ r.l.u. along $(0,k,0)$.
    \label{figs:perp}}
\end{figure}

\section{Interchain couplings}

Figure~\ref{figs:perp}(a) presents the INS spectra along the $(0,k,0)$ direction. Data is integrated in the range of $\delta h = [-1.2, 1.2]$ r.l.u. along $(h,0,h)$. The integration range along the $(h,0,-h)$ direction is selected as $\delta h = [0.2, 0.3]$ r.l.u. where the scattering intensity is the strongest at the top of the lower boundary of the spinon continuum (see Fig.~2b of the main text). No dispersion is observed within the instrumental energy resolution, which is consistent with the relatively large separation of $\sim 8.14$~\AA\ between the V-O layers along the $b$ axis.

Figure~\ref{figs:perp}(b) presents the INS spectra along the $(h,0,h)$ direction. Data is integrated in the range of $\delta k = [-4, 4]$ r.l.u. along $(0,k,0)$. The integration range along the $(h,0,-h)$ direction is selected as $\delta h = [0.7, 0.8]$ r.l.u. for better coverage of the detected area. A weak dispersion with a bandwidth of $\sim 0.5$~meV is observed, which is consistent with the existence of a magnetic long range order transition at $T_N = 5.2$~K.

\section{Instrumental energy resolution}
Detailed information for the instrumental energy resolution of SEQUOIA can be found in the instrument webpage (https://neutrons.ornl.gov/sequoia/users). For the setups used in our experiments, the instrumental energy resolution can be described as a function of the energy transfer $E$:
\begin{align}
    \delta E|_{E_i = 35~\rm{meV}} &= 2.5913\times 10^{-6} \times E^3 +4.1543\times 10^{-4} \times E^2 - 0.0454 \times E +1.5881 \\
    \delta E|_{E_i = 10~\rm{meV}} &= 8.7056\times 10^{-6} \times E^3 +5.1225\times 10^{-4} \times E^2 - 0.0147 \times E +0.2104 \\
    \delta E|_{E_i = 4~\rm{meV}} &= 5.4111\times 10^{-5} \times E^3 +1.1827\times 10^{-3} \times E^2 - 0.0136 \times E +0.0695
 \end{align}

\begin{figure}[t!]
    \includegraphics[width=0.55\textwidth]{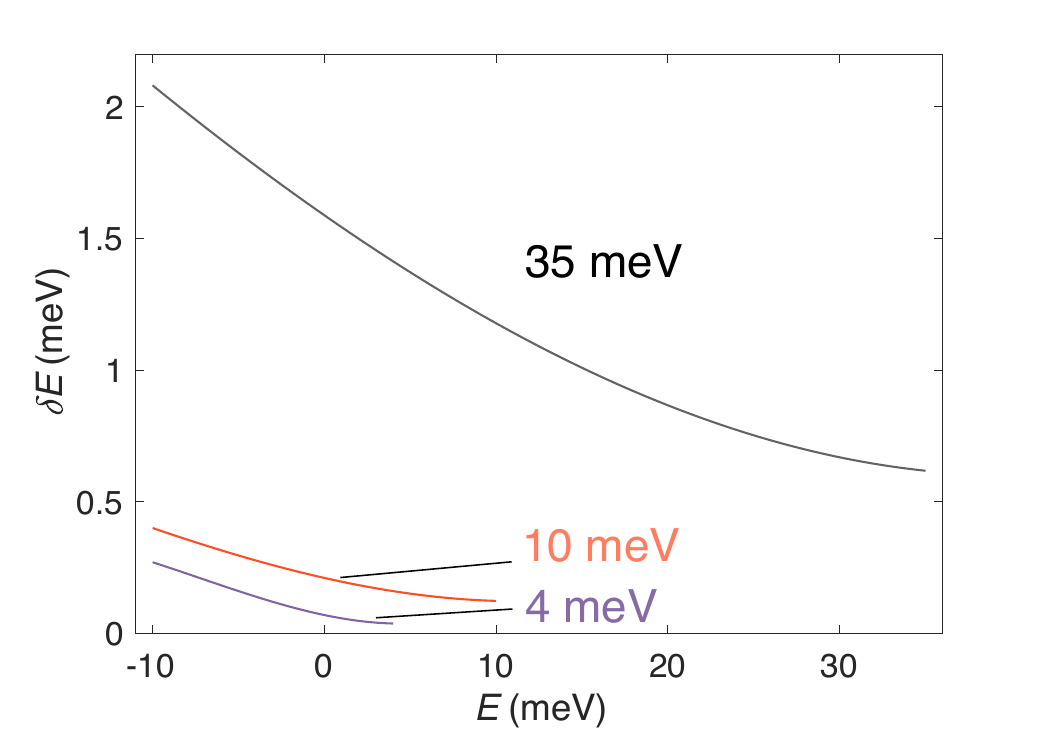}
    \caption{Instrumental energy resolution on SEQUOIA for $E_i = 35$, 10, and 4~meV. The chopper information for each $E_i$ is listed in the main text.
    \label{figs:res}}
\end{figure}

 As an example, we plot the instrumental energy resolution for $E_i = 35$~meV in Fig.~\ref{figs:res}. At the elastic line of $E = 0$, the energy resolution is $\delta E = 1.59$~meV. The energy resolution improves at higher energy transfers, which is a general dependence for the energy resolution of direct geometry time-of-flight neutron spectrometers.

\section{REPRODUCING THE DMRG RESULTS.}
\subsection{Install DMRG++}
Here, detailed instructions are provided to reproduce the DMRG results presented in the main text. We closely follow the description in the supplemental material of Ref. \cite{Scheie2021Witnessing}, where similar calculations were performed. The results used in this work were obtained with the open-source DMRG++ computer software~\cite{alvarez2009density} with version 6.05 and PsimagLite version 3.04. DMRG++ is free and open to community contributions. Please find more details at \url{https://github.com/g1257/dmrgpp}.

The DMRG++ computer program~\cite{alvarez2009density} can be obtained with:
\texttt{
\\ \hspace*{\fill} \\
git clone https://github.com/g1257/dmrgpp.git
}
\\ \hspace*{\fill} \\
and PsimagLite:
\texttt{
\\ \hspace*{\fill} \\
git clone https://github.com/g1257/PsimagLite.git
}
\\ \hspace*{\fill} \\
To compile, use:
\texttt{
\\ \hspace*{\fill} \\
cd PsimagLite/lib
\\perl configure.pl
\\make
\\ \hspace*{\fill} \\
cd ../../dmrgpp/src
\\perl configure.pl
\\make
\\ \hspace*{\fill} \\
}
For brevity we also run the following commands to define the environment variables:
\texttt{
\\ \hspace*{\fill} \\
export PATH="<PATH-TO-DMRG++>/src:\$PATH"\\
export SCRIPTS="<PATH-TO-DMRG++>/scripts"
\\ \hspace*{\fill} \\
}
The documentation can be found at \url{https://g1257.github.io/dmrgPlusPlus/manual.html} or obtained by running:
\texttt{
\\ \hspace*{\fill} \\
cd ../..\\
git clone https://github.com/g1257/thesis.git\\
cd dmrgpp/doc\\
ln -s ../../thesis/thesis.bib\\
make manual.pdf
\\ \hspace*{\fill} \\
}

\subsection{Obtaining zero-temperature spectra}
The main instructions to reproduce our $T = 0$ results are described as follows. First, to obtain the ground state, run the DMRG++ package with an input file by using \texttt{dmrg -f inputGS.ain}. The input file \texttt{inputGS.ain} has the form as below:
\texttt{
\\ \hspace*{\fill} \\
\#\#Ainur1.0\\
TotalNumberOfSites=100;\\
NumberOfTerms=2;
\\ \hspace*{\fill} \\
\#\#\# $1/2(S^+S^- + S^-S^+)$ part\\
gt0:DegreesOfFreedom=1;\\
gt0:GeometryKind="chain";\\
gt0:GeometryOptions="ConstantValues";\\
gt0:dir0:Connectors=[1.0];
\\ \hspace*{\fill} \\
\#\#\# $S^zS^z$ part\\
gt1:DegreesOfFreedom=1;\\
gt1:GeometryKind="chain";\\
gt1:GeometryOptions="ConstantValues";\\
gt1:dir0:Connectors=[1.0];
\\ \hspace*{\fill} \\
Model="Heisenberg";\\
HeisenbergTwiceS=1;
\\ \hspace*{\fill} \\
SolverOptions="twositedmrg,calcAndPrintEntropies";\\
Version="version";\\
OutputFile="dataGS";\\
InfiniteLoopKeptStates=1000;\\
FiniteLoops=[\\
{[ 49, 1000, 0],}\\
{[-98, 1000, 0],}\\
{[ 98, 1000, 0],}\\
{[-98, 1000, 0],}\\
{[ 98, 1000, 1]];}
\\ \hspace*{\fill} \\
\# Keep a maximum of 1000 states, but allow\\
\# truncation with tolerance and minimum states\\
TruncationTolerance="1e-10,100";
\\ \hspace*{\fill} \\
\# Tolerance for Lanczos\\
LanczosEps=1e-10;\\
int LanczosSteps=600;
\\ \hspace*{\fill} \\
Threads=4;\\
TargetSzPlusConst=50;
\\ \hspace*{\fill} \\
}

Note that the input is for $S = 1/2$. If included, the parameter \texttt{TargetSzPlusConst} should equal $S^z + S \cdot N$, where $S^z$ is the targeted $S^z$ sector and $N$ is the size of the chain.

The next step is to calculate dynamics in a subdirectory \texttt{Szz}. Since the Heisenberg model is isotropic it is sufficient to consider only the $S^{zz}(q,\omega)$ component. Note that we restart from the above saved ground state and the input file \texttt{inputSzz.ado} is given by:
\texttt{
\\ \hspace*{\fill} \\
\#\#Ainur1.0\\
TotalNumberOfSites=100;\\
NumberOfTerms=2;
\\ \hspace*{\fill} \\
\#\#\# $1/2(S^+S^- + S^-S^+)$ part\\
gt0:DegreesOfFreedom=1;\\
gt0:GeometryKind="chain";\\
gt0:GeometryOptions="ConstantValues";\\
gt0:dir0:Connectors=[1.0];
\\ \hspace*{\fill} \\
\#\#\# $S^zS^z$ part\\
gt1:DegreesOfFreedom=1;\\
gt1:GeometryKind="chain";\\
gt1:GeometryOptions="ConstantValues";\\
gt1:dir0:Connectors=[1.0];
\\ \hspace*{\fill} \\
Model="Heisenberg";\\
HeisenbergTwiceS=1;
\\ \hspace*{\fill} \\
SolverOptions="twositedmrg,restart,minimizeDisk,CorrectionVectorTargeting";\\
Version="version";\\
\# RestartFilename is the name of the GS .hd5 file (extension is not needed)\\
RestartFilename="../dataGS";\\
InfiniteLoopKeptStates=1000;\\
\#\#\# The finite loops pick up where gs run ended! I.e. the edge.\\
FiniteLoops=[\\
{[-98, 1000, 2],}\\
{[98, 1000, 2]];}
\\ \hspace*{\fill} \\
\# Keep a maximum of 1000 states, but allow\\
\# truncation with tolerance and minimum states\\
TruncationTolerance="1e-10,100";
\\ \hspace*{\fill} \\
\# Tolerance for Lanczos\\
LanczosEps=1e-10;\\
int LanczosSteps=600;\\
Threads=4;\\
TargetSzPlusConst=50;
\\ \hspace*{\fill} \\
\# The weight of the g.s. in the density matrix\\
GsWeight=0.1;\\
\# Legacy thing, set to 0\\
CorrectionA=0;\\
\# Fermion spectra has sign changes in denominator. For boson operators (as in here) set it to 0\\
DynamicDmrgType=0;\\
\# The site(s) where to apply the operator below. Here it is the center site.\\
TSPSites=[50];\\
\# The delay in loop units before applying the operator. Set to 0 for all restarts to avoid delays.\\
TSPLoops=[0];\\
\# If more than one operator is to be applied, how they should be combined.\\
\# Irrelevant if only one operator is applied, as is the case here.\\
TSPProductOrSum="sum";\\
\# How the operator to be applied will be specified\\
string TSPOp0:TSPOperator="expression";\\
\# The operator expression\\
string TSPOp0:OperatorExpression="sz";\\
\# How is the freq. given in the denominator (Matsubara is the other option)\\
CorrectionVectorFreqType="Real";\\
\# This is a dollarized input, so the omega will change from input to input.\\
CorrectionVectorOmega=\$omega;\\
\# The broadening for the spectrum in omega + i*eta\\
CorrectionVectorEta=0.10;\\
\# The algorithm\\
CorrectionVectorAlgorithm="Krylov";
\\ \hspace*{\fill} \\
\#The labels below are ONLY read by manyOmegas.pl script\\
\# How many inputs files to create\\
\#OmegaTotal=200\\
\# Which one is the first omega value\\
\#OmegaBegin=0.0\\
\# Which is the "step" in omega\\
\#OmegaStep=0.025\\
\# Because the script will also be creating the batches, indicate what to measure in the batches\\
\#Observable=sz
\\ \hspace*{\fill} \\
}

In the correction vector approach, each $\omega$ will correspond to one input.
Here, 200 inputs, corresponding to 200 $\omega$'s, can be obtained and submitted by using the \texttt{manyOmegas.pl script}:
\texttt{
\\ \hspace*{\fill} \\
perl -I \$\{SCRIPTS\} \$\{SCRIPTS\}/manyOmegas.pl inputSzz.ado BatchTemplate <test/submit>.
\\ \hspace*{\fill} \\
}
Before submitting with many $\omega$'s, it is recommended to run with test first or submit with few $\omega$'s to verify correctness, because those calculations can be time- and computer-consuming. The BatchTemplate should be a suitable job submission script, such as a PBS script, depending on the machine and scheduler. Note that it must contain the following line:
\texttt{
\\ \hspace*{\fill} \\
dmrg -f \$\$input "<X0|\$\$obs|P1>,<X0|\$\$obs|P2>,<X0|\$\$obs|P3>" -p 10
\\ \hspace*{\fill} \\
}
which allows \texttt{manyOmegas.pl} to reference the appropriate input in each job batch. 
After all outputs have been automatically generated and no errors shown in ``error.log", we can use
\texttt{
\\ \hspace*{\fill} \\
perl -I \$\{SCRIPTS\} \$\{SCRIPTS\}/procOmegas.pl -f inputSzz.ado -p\\
perl \$\{SCRIPTS\}/pgfplot.pl
\\ \hspace*{\fill} \\
}
to process and plot the DMRG results.

\subsection{Obtaining finite-temperature spectra}
Next, to obtain the results of the $T > 0$ calculation, we proceed in three steps as described in the following. First, the $T = \infty $ state is obtained as the ground state of a fictitious ``entangler'' Hamiltonian $H_{\rm E}$ acting in an enlarged Hilbert space \cite{Feiguin2005finite,Feiguin2010Spectral,Nocera2016symmetry}. Second, the physical system is cooled through evolving in imaginary time with the physical Hamiltonian $H$ acting only on physical sites, i.e. we evolve with $H  \otimes I$, where $I$ is the identity operator in the ancilla space. Third, the operator $H  \otimes I + I \otimes (-H)$ is adopted to calculate dynamics. We note that the  $T > 0$ DMRG results discussed in the main text are very time consuming. 

In the first step, a conventional (grand canonical) entangler is used in our DMRG calculations, such that the enlarged system (physical and ancilla sites) can be regarded as a two-leg spin ladder system, with physical sites on one leg (with even sites 0, 2, 4,...) and ancilla sites on the other (with odd sites 1, 3, 5,...). The entangler Hamiltonian is chosen so that its ground state corresponds to the $T = \infty $  state of the physical system.
By running \texttt{dmrg -f Entangler.ain}, the ground state of $H_{\rm E}$ can be obtained.  The \texttt{Entangler.ain} is given as:
\texttt{
\\ \hspace*{\fill} \\
\#\#Ainur1.0\\
TotalNumberOfSites=100;\\
NumberOfTerms=2;
\\ \hspace*{\fill} \\
\#\#\# $1/2(S^+S^- + S^-S^+)$ part\\
gt0:DegreesOfFreedom=1;\\
gt0:GeometryKind="ladder";\\
gt0:LadderLeg=2;\\
gt0:GeometryOptions="ConstantValues";\\
gt0:dir0:Connectors=[0.0];\\
gt0:dir1:Connectors=[-10.0];\\
integer gt0:IsPeriodicX=0;
\\ \hspace*{\fill} \\
\#\#\# $S^zS^z$ part\\
gt1:DegreesOfFreedom=1;\\
gt1:GeometryKind="chain";\\
gt1:GeometryOptions="ConstantValues";\\
gt1:dir0:Connectors=[0];\\
integer gt1:IsPeriodicX=0;
\\ \hspace*{\fill} \\
Model="Heisenberg";\\
HeisenbergTwiceS=1;
\\ \hspace*{\fill} \\
SolverOptions="twositedmrg,MatrixVectorOnTheFly";\\
Version="version";\\
InfiniteLoopKeptStates=1000;\\
FiniteLoops=[\\
{[ 49, 1000, 0],}\\
{[-98, 1000, 0],}\\
{[ 98, 1000, 0],}\\
{[-98, 1000, 0],}\\
{[ 98, 1000, 0]];}\\
\# Keep a maximum of 1000 states, but allow\\
\# truncation with tolerance and minimum states\\
TruncationTolerance="1e-10,100";\\
\# Tolerance for Lanczos\\
LanczosEps=1e-8;\\
int LanczosSteps=250;\\
Threads=4;
\\ \hspace*{\fill} \\
}

\texttt{GeometryKind="Ladder"} for \texttt{gt1} can be used instead of \texttt{gt1:GeometryKind="chain"}. Next, we make a subdirectory \texttt{evolution1} and start the imaginary time evolution by running \texttt{dmrg -f evolution1.ain}. 
The \texttt{evolution1.ain} file is shown below:
\texttt{
\\ \hspace*{\fill} \\
\#\#Ainur1.0\\
TotalNumberOfSites=100;\\
NumberOfTerms=2;
\\ \hspace*{\fill} \\
\#\#\# $1/2(S^+S^- + S^-S^+)$ part\\
gt0:DegreesOfFreedom=1;\\
gt0:GeometryKind="ladder";\\
gt0:LadderLeg=2;\\
gt0:GeometryOptions="none";\\
gt0:dir0:Connectors=[1.0, 0.0, 1.0, 0.0, 1.0, 0.0, 1.0, 0.0, 1.0, 0.0, 1.0, 0.0, 1.0, 0.0, 1.0, 0.0, 1.0, 0.0, 1.0, 0.0, 1.0, 0.0, 1.0, 0.0, 1.0, 0.0, 1.0, 0.0, 1.0, 0.0, 1.0, 0.0, 1.0, 0.0, 1.0, 0.0, 1.0, 0.0, 1.0, 0.0, 1.0, 0.0, 1.0, 0.0, 1.0, 0.0, 1.0, 0.0, 1.0, 0.0, 1.0, 0.0, 1.0, 0.0, 1.0, 0.0, 1.0, 0.0, 1.0, 0.0, 1.0, 0.0, 1.0, 0.0, 1.0, 0.0, 1.0, 0.0, 1.0, 0.0, 1.0, 0.0, 1.0, 0.0, 1.0, 0.0, 1.0, 0.0, 1.0, 0.0, 1.0, 0.0, 1.0, 0.0, 1.0, 0.0, 1.0, 0.0, 1.0, 0.0, 1.0, 0.0, 1.0, 0.0, 1.0, 0.0, 1.0, 0.0];\\
gt0:dir1:Connectors=[0.0, 0.0, 0.0, 0.0, 0.0, 0.0, 0.0, 0.0, 0.0, 0.0, 0.0, 0.0, 0.0, 0.0, 0.0, 0.0, 0.0, 0.0, 0.0, 0.0, 0.0, 0.0, 0.0, 0.0, 0.0, 0.0, 0.0, 0.0, 0.0, 0.0, 0.0, 0.0, 0.0, 0.0, 0.0, 0.0, 0.0, 0.0, 0.0, 0.0, 0.0, 0.0, 0.0, 0.0, 0.0, 0.0, 0.0, 0.0, 0.0, 0.0];\\
integer gt0:IsPeriodicX=0;
\\ \hspace*{\fill} \\
\#\#\# $S^zS^z$ part\\
gt1:DegreesOfFreedom=1;\\
gt1:GeometryKind="ladder";\\
gt1:LadderLeg=2;\\
gt1:GeometryOptions="none";\\
gt1:dir0:Connectors=[1.0, 0.0, 1.0, 0.0, 1.0, 0.0, 1.0, 0.0, 1.0, 0.0, 1.0, 0.0, 1.0, 0.0, 1.0, 0.0, 1.0, 0.0, 1.0, 0.0, 1.0, 0.0, 1.0, 0.0, 1.0, 0.0, 1.0, 0.0, 1.0, 0.0, 1.0, 0.0, 1.0, 0.0, 1.0, 0.0, 1.0, 0.0, 1.0, 0.0, 1.0, 0.0, 1.0, 0.0, 1.0, 0.0, 1.0, 0.0, 1.0, 0.0, 1.0, 0.0, 1.0, 0.0, 1.0, 0.0, 1.0, 0.0, 1.0, 0.0, 1.0, 0.0, 1.0, 0.0, 1.0, 0.0, 1.0, 0.0, 1.0, 0.0, 1.0, 0.0, 1.0, 0.0, 1.0, 0.0, 1.0, 0.0, 1.0, 0.0, 1.0, 0.0, 1.0, 0.0, 1.0, 0.0, 1.0, 0.0, 1.0, 0.0, 1.0, 0.0, 1.0, 0.0, 1.0, 0.0, 1.0, 0.0];\\
gt1:dir1:Connectors=[0.0, 0.0, 0.0, 0.0, 0.0, 0.0, 0.0, 0.0, 0.0, 0.0, 0.0, 0.0, 0.0, 0.0, 0.0, 0.0, 0.0, 0.0, 0.0, 0.0, 0.0, 0.0, 0.0, 0.0, 0.0, 0.0, 0.0, 0.0, 0.0, 0.0, 0.0, 0.0, 0.0, 0.0, 0.0, 0.0, 0.0, 0.0, 0.0, 0.0, 0.0, 0.0, 0.0, 0.0, 0.0, 0.0, 0.0, 0.0, 0.0, 0.0];\\
integer gt1:IsPeriodicX=0;
\\ \hspace*{\fill} \\
Model="Heisenberg";\\
HeisenbergTwiceS=1;
\\ \hspace*{\fill} \\
string PrintHamiltonianAverage="s==c";\\
string RecoverySave="@M=100,@keep,1==1";
\\ \hspace*{\fill} \\
SolverOptions="twositedmrg,restart,TargetingAncilla";\\
Version="version";\\
InfiniteLoopKeptStates=1000;\\
FiniteLoops=[\\
{[-98, 1000, 2],}\\
{[ 98, 1000, 2]];}\\
RepeatFiniteLoopsTimes=21;\\
\# Keep a maximum of 1000 states, but allow\\
\# truncation with tolerance and minimum states\\
TruncationTolerance="1e-10,100";\\
\# Tolerance for Lanczos\\
LanczosEps=1e-8;\\
int LanczosSteps=250;\\
Threads=4;
\\ \hspace*{\fill} \\
RestartFilename="../entangler";\\
TSPTau=0.1;\\
TSPTimeSteps=5;\\
TSPAdvanceEach=98;\\
TSPAlgorithm="Krylov";\\
TSPSites=[50];\\
TSPLoops=[0];\\
TSPProductOrSum="sum";\\
GsWeight=0.1;
\\ \hspace*{\fill} \\
string TSPOp0:TSPOperator="expression";\\
string TSPOp0:OperatorExpression="identity";
\\ \hspace*{\fill} \\
}

Since we only act on physical sites, the \texttt{gt?:dir1:Connectors} should all be zero, as they describe couplings across rungs. There are  50 ($N/2$)  such rungs, and a value for each needs to be listed in the array in the DMRG input file. The \texttt{gt?:dir0:Connectors} are used to describe couplings along legs, starting at site $j$ corresponding to the $j$th position in the array (indexed from zero).
Note that every other leg coupling is set to zero to only couple physical sites.  For open boundary conditions there are $N - 2$ such bonds that need to be included in the array in the input.

In addition, we use the Krylov imaginary time evolution with a time step $\tau= 0.1$ in our DMRG calculations. 
The time evolution is done with an evolution operator $\exp[-H\tau]$. The evolved state is identified with $\left| \psi (\beta)\right\rangle = \exp[ -\frac{\beta H}{2}] \left| \psi(0)\right\rangle$, where $\left| \psi(0)\right\rangle$ is the ground state of $H_E$ and $\beta$ is the inverse temperature \cite{Feiguin2005finite}.
Hence, each $\tau$ corresponds to $\beta'=\beta/2$. 
Using \texttt{h5dump -d /Def/FinalPsi/TimeSerializer/Time <hd5>}, the imaginary time $\beta'$ can be obtained, where \texttt{<hd5>} should be replaced with the name of the \texttt{hd5} file of interest.

To get the imaginary times corresponding to the values of experimental temperatures, we did additional restarts from appropriate \texttt{hd5} files output in the main temperature evolution loop, while tuning the $\tau$ step \texttt{TSPTau}. Specifically, the targeted dimensionless $\beta'$ value can be obtained as $\beta'(J,T)= J/(2k_BT)$, where $k_{\rm B}=0.08617333262$ meV/K, $J$ and $T$ are given in meV and Kelvin, respectively. Furthermore, the arguments to \texttt{RecoverySave} indicate that we kept a maximum of \texttt{100 hd5} outputs, and output one in every loop (when the condition 1 == 1 holds). The number of \texttt{hd5} files is either the \texttt{RecoverySave} maximum or the total number of finite loops, whichever is smaller. In this input, there are a total of 44 finite loops, so that we got 44 \texttt{hd5} files. Here we take $J=7.1$ meV and $T=50$ K as an example, where the targeted $\beta'=0.824$ with 3 significant digits. After the evolution with a time step $\tau= 0.1$, we get $\beta'=0.8$ in \texttt{Recovery8evolution1.hd5}. Then restarting from \texttt{Recovery8evolution1.hd5} with $\tau= 0.012$, we will get the targeted $\beta'=0.824$ in \texttt{Recovery2evolution2.hd5} as in the new evolution loop. In evolution2, we make a subdirectory \texttt{evolution2} and \texttt{cp evolution1.ain evolution2/evolution2.ain}. Then, modify the following parameters in \texttt{evolution2.ain}.
\texttt{
\\ \hspace*{\fill} \\
RepeatFiniteLoopsTimes=2; \\
RestartFilename="../Recovery8evolution1"; \\
TSPTau=0.012;
\\ \hspace*{\fill} \\
}

Finally, the dynamics calculation proceeds similarly to the $T = 0$ case, but with number of sites and precision as in the preceding $T > 0$ step.  We make a subdirectory \texttt{sqw} and prepare \texttt{sqw.ain} as follows
\texttt{
\\ \hspace*{\fill} \\
\#\#Ainur1.0\\
TotalNumberOfSites=100;\\
NumberOfTerms=2;
\\ \hspace*{\fill} \\
string GeometrySubKind="GrandCanonical";
\\ \hspace*{\fill} \\
\#\#\# $1/2(S^+S^- + S^-S^+)$ part\\
gt0:DegreesOfFreedom=1;\\
gt0:GeometryKind="ladder";\\
gt0:LadderLeg=2;\\
gt0:GeometryOptions="none";\\
gt0:dir0:Connectors=[1.0, -1.0, 1.0, -1.0, 1.0, -1.0, 1.0, -1.0, 1.0, -1.0, 1.0, -1.0, 1.0, -1.0, 1.0, -1.0, 1.0, -1.0, 1.0, -1.0, 1.0, -1.0, 1.0, -1.0, 1.0, -1.0, 1.0, -1.0, 1.0, -1.0, 1.0, -1.0, 1.0, -1.0, 1.0, -1.0, 1.0, -1.0, 1.0, -1.0, 1.0, -1.0, 1.0, -1.0, 1.0, -1.0, 1.0, -1.0, 1.0, -1.0, 1.0, -1.0, 1.0, -1.0, 1.0, -1.0, 1.0, -1.0, 1.0, -1.0, 1.0, -1.0, 1.0, -1.0, 1.0, -1.0, 1.0, -1.0, 1.0, -1.0, 1.0, -1.0, 1.0, -1.0, 1.0, -1.0, 1.0, -1.0, 1.0, -1.0, 1.0, -1.0, 1.0, -1.0, 1.0, -1.0, 1.0, -1.0, 1.0, -1.0, 1.0, -1.0, 1.0, -1.0, 1.0, -1.0, 1.0, -1.0];\\
gt0:dir1:Connectors=[0.0, 0.0, 0.0, 0.0, 0.0, 0.0, 0.0, 0.0, 0.0, 0.0, 0.0, 0.0, 0.0, 0.0, 0.0, 0.0, 0.0, 0.0, 0.0, 0.0, 0.0, 0.0, 0.0, 0.0, 0.0, 0.0, 0.0, 0.0, 0.0, 0.0, 0.0, 0.0, 0.0, 0.0, 0.0, 0.0, 0.0, 0.0, 0.0, 0.0, 0.0, 0.0, 0.0, 0.0, 0.0, 0.0, 0.0, 0.0, 0.0, 0.0];
\\ \hspace*{\fill} \\
\#\#\# $S^zS^z$ part\\
gt1:DegreesOfFreedom=1;\\
gt1:GeometryKind="ladder";\\
gt1:LadderLeg=2;\\
gt1:GeometryOptions="none";\\
gt1:dir0:Connectors=[1.0, -1.0, 1.0, -1.0, 1.0, -1.0, 1.0, -1.0, 1.0, -1.0, 1.0, -1.0, 1.0, -1.0, 1.0, -1.0, 1.0, -1.0, 1.0, -1.0, 1.0, -1.0, 1.0, -1.0, 1.0, -1.0, 1.0, -1.0, 1.0, -1.0, 1.0, -1.0, 1.0, -1.0, 1.0, -1.0, 1.0, -1.0, 1.0, -1.0, 1.0, -1.0, 1.0, -1.0, 1.0, -1.0, 1.0, -1.0, 1.0, -1.0, 1.0, -1.0, 1.0, -1.0, 1.0, -1.0, 1.0, -1.0, 1.0, -1.0, 1.0, -1.0, 1.0, -1.0, 1.0, -1.0, 1.0, -1.0, 1.0, -1.0, 1.0, -1.0, 1.0, -1.0, 1.0, -1.0, 1.0, -1.0, 1.0, -1.0, 1.0, -1.0, 1.0, -1.0, 1.0, -1.0, 1.0, -1.0, 1.0, -1.0, 1.0, -1.0, 1.0, -1.0, 1.0, -1.0, 1.0, -1.0];\\
gt1:dir1:Connectors=[0.0, 0.0, 0.0, 0.0, 0.0, 0.0, 0.0, 0.0, 0.0, 0.0, 0.0, 0.0, 0.0, 0.0, 0.0, 0.0, 0.0, 0.0, 0.0, 0.0, 0.0, 0.0, 0.0, 0.0, 0.0, 0.0, 0.0, 0.0, 0.0, 0.0, 0.0, 0.0, 0.0, 0.0, 0.0, 0.0, 0.0, 0.0, 0.0, 0.0, 0.0, 0.0, 0.0, 0.0, 0.0, 0.0, 0.0, 0.0, 0.0, 0.0];
\\ \hspace*{\fill} \\
Model="Heisenberg";\\
HeisenbergTwiceS=1;
\\ \hspace*{\fill} \\
SolverOptions="CorrectionVectorTargeting,restart,twositedmrg,minimizeDisk,fixLegacyBugs";\\
Version="version";\\
\# RestartFilename is the name of the GS .hd5 file (extension is not needed)\\
RestartFilename="../Recovery2evolution2";\\
InfiniteLoopKeptStates=1000;\\
\#\#\# The finite loops pick up where gs run ended! I.e. the edge.\\
FiniteLoops=[\\
{[-98, 1000, 2],}\\
{[98, 1000, 2]];}\\
\# Keep a maximum of 1000 states, but allow\\
\# truncation with tolerance and minimum states\\
TruncationTolerance="1e-10,100";\\
\# Tolerance for Lanczos\\
LanczosEps=1e-8;\\
int LanczosSteps=250;\\
Threads=4;\\
\#TargetSzPlusConst=50;\\
\\ \hspace*{\fill} \\
integer RestartSourceTvForPsi=0;\\
vector RestartMappingTvs=[-1, -1, -1, -1];\\
integer RestartMapStages=0;\\
integer TridiagSteps=400;\\
real TridiagEps=1e-9;\\
\\ \hspace*{\fill} \\
\# The weight of the g.s. in the density matrix\\
GsWeight=0.1;\\
\# Legacy thing, set to 0\\
CorrectionA=0;\\
\# Fermion spectra has sign changes in denominator. For boson operators (as in here) set it to 0\\
DynamicDmrgType=0;\\
\# The site(s) where to apply the operator below. Here it is the center site.\\
TSPSites=[50];\\
\# The delay in loop units before applying the operator. Set to 0 for all restarts to avoid delays.\\
TSPLoops=[0];\\
\# If more than one operator is to be applied, how they should be combined.\\
\# Irrelevant if only one operator is applied, as is the case here.\\
TSPProductOrSum="sum";\\
\# How the operator to be applied will be specified\\
string TSPOp0:TSPOperator="expression";\\
\# The operator expression\\
string TSPOp0:OperatorExpression="sz";\\
\#When this line is not there, DMRG++ assumes $\omega$ is measured relative to the ground state energy. This is usually fine, but for finite-T spectra it assumes it should use the ground state energy of the entangler Hamiltonian.\\
real TSPEnergyForExp=0;\\
\# How is the freq. given in the denominator (Matsubara is the other option)\\
CorrectionVectorFreqType="Real";\\
\# This is a dollarized input, so the omega will change from input to input.\\
CorrectionVectorOmega=\$omega;\\
\# The broadening for the spectrum in omega + i*eta\\
CorrectionVectorEta=0.10;\\
\# The algorithm\\
CorrectionVectorAlgorithm="Krylov";\\
\#The labels below are ONLY read by manyOmegas.pl script\\
\# How many inputs files to create\\
\#OmegaTotal=200\\
\# Which one is the first omega value\\
\#OmegaBegin=0.0\\
\# Which is the "step" in omega\\
\#OmegaStep=0.025\\
\# Because the script will also be creating the batches, indicate what to measure in the batches\\
\#Observable=sz
\\ \hspace*{\fill} \\
}

The restart filename should be chosen to match the \texttt{hd5} file of interest. Note here that we calculate dynamics with $H  \otimes I + I \otimes (-H)$, where $H  \otimes I$ acts nontrivially only on physical sites and $I \otimes (-H)$ acts nontrivially only on ancilla sites. All rung couplings are zero.

Because we used $J=1$ throughout the calculations, we need to scale the obtained data $\omega$ to $\omega*7.1$ and $S(q,\omega)$ to $S(q,\omega)/7.1$, respectively, to match the Sr$_2$V$_3$O$_9$ energy scale.


\begin{thebibliography}{47}%
  \makeatletter
  \providecommand \@ifxundefined [1]{%
   \@ifx{#1\undefined}
  }%
  \providecommand \@ifnum [1]{%
   \ifnum #1\expandafter \@firstoftwo
   \else \expandafter \@secondoftwo
   \fi
  }%
  \providecommand \@ifx [1]{%
   \ifx #1\expandafter \@firstoftwo
   \else \expandafter \@secondoftwo
   \fi
  }%
  \providecommand \natexlab [1]{#1}%
  \providecommand \enquote  [1]{``#1''}%
  \providecommand \bibnamefont  [1]{#1}%
  \providecommand \bibfnamefont [1]{#1}%
  \providecommand \citenamefont [1]{#1}%
  \providecommand \href@noop [0]{\@secondoftwo}%
  \providecommand \href [0]{\begingroup \@sanitize@url \@href}%
  \providecommand \@href[1]{\@@startlink{#1}\@@href}%
  \providecommand \@@href[1]{\endgroup#1\@@endlink}%
  \providecommand \@sanitize@url [0]{\catcode `\\12\catcode `\$12\catcode
    `\&12\catcode `\#12\catcode `\^12\catcode `\_12\catcode `\%12\relax}%
  \providecommand \@@startlink[1]{}%
  \providecommand \@@endlink[0]{}%
  \providecommand \url  [0]{\begingroup\@sanitize@url \@url }%
  \providecommand \@url [1]{\endgroup\@href {#1}{\urlprefix }}%
  \providecommand \urlprefix  [0]{URL }%
  \providecommand \Eprint [0]{\href }%
  \providecommand \doibase [0]{https://doi.org/}%
  \providecommand \selectlanguage [0]{\@gobble}%
  \providecommand \bibinfo  [0]{\@secondoftwo}%
  \providecommand \bibfield  [0]{\@secondoftwo}%
  \providecommand \translation [1]{[#1]}%
  \providecommand \BibitemOpen [0]{}%
  \providecommand \bibitemStop [0]{}%
  \providecommand \bibitemNoStop [0]{.\EOS\space}%
  \providecommand \EOS [0]{\spacefactor3000\relax}%
  \providecommand \BibitemShut  [1]{\csname bibitem#1\endcsname}%
  \let\auto@bib@innerbib\@empty
  \bibitem [{\citenamefont {Giamarchi}(2003)}]{giamarchi_quantum_2003}%
    \BibitemOpen
    \bibfield  {author} {\bibinfo {author} {\bibfnamefont {T.}~\bibnamefont
    {Giamarchi}},\ }\href
    {https://global.oup.com/academic/product/quantum-physics-in-one-dimension-9780198525004?q=Quantum%20Physics%20in%20One%20Dimension&lang=en&cc=us}
    {\emph {\bibinfo {title} {Quantum Physics in One Dimension}}}\ (\bibinfo
    {publisher} {Clarendon Press, Oxford},\ \bibinfo {year} {2003})\BibitemShut
    {NoStop}%
  \bibitem [{\citenamefont {Bethe}(1931)}]{bethe_zur_1931}%
    \BibitemOpen
    \bibfield  {author} {\bibinfo {author} {\bibfnamefont {H.}~\bibnamefont
    {Bethe}},\ }\bibfield  {title} {\bibinfo {title} {Zur {Theorie} der
    {Metalle}. {I}. {Eigenwerte} und {Eigenfunktionen} der linearen
    {Atomkette}},\ }\href {https://doi.org/10.1007/BF01341708} {\bibfield
    {journal} {\bibinfo  {journal} {Z. Phys.}\ }\textbf {\bibinfo {volume}
    {71}},\ \bibinfo {pages} {205} (\bibinfo {year} {1931})}\BibitemShut
    {NoStop}%
  \bibitem [{\citenamefont {Faddeev}\ and\ \citenamefont
    {Takhtajan}(1981)}]{faddeev_what_1981}%
    \BibitemOpen
    \bibfield  {author} {\bibinfo {author} {\bibfnamefont {L.}~\bibnamefont
    {Faddeev}}\ and\ \bibinfo {author} {\bibfnamefont {L.}~\bibnamefont
    {Takhtajan}},\ }\bibfield  {title} {\bibinfo {title} {What is the spin of a
    spin wave?},\ }\href
    {https://doi.org/https://doi.org/10.1016/0375-9601(81)90335-2} {\bibfield
    {journal} {\bibinfo  {journal} {Physics Letters A}\ }\textbf {\bibinfo
    {volume} {85}},\ \bibinfo {pages} {375} (\bibinfo {year} {1981})}\BibitemShut
    {NoStop}%
  \bibitem [{\citenamefont {Affleck}\ \emph {et~al.}(1987)\citenamefont
    {Affleck}, \citenamefont {Kennedy}, \citenamefont {Lieb},\ and\ \citenamefont
    {Tasaki}}]{affleck_rigorous_1987}%
    \BibitemOpen
    \bibfield  {author} {\bibinfo {author} {\bibfnamefont {I.}~\bibnamefont
    {Affleck}}, \bibinfo {author} {\bibfnamefont {T.}~\bibnamefont {Kennedy}},
    \bibinfo {author} {\bibfnamefont {E.~H.}\ \bibnamefont {Lieb}},\ and\
    \bibinfo {author} {\bibfnamefont {H.}~\bibnamefont {Tasaki}},\ }\bibfield
    {title} {\bibinfo {title} {Rigorous results on valence-bond ground states in
    antiferromagnets},\ }\href
    {http://link.aps.org/doi/10.1103/PhysRevLett.59.799} {\bibfield  {journal}
    {\bibinfo  {journal} {Phys. Rev. Lett.}\ }\textbf {\bibinfo {volume} {59}},\
    \bibinfo {pages} {799} (\bibinfo {year} {1987})}\BibitemShut {NoStop}%
  \bibitem [{\citenamefont {Sutherland}(2004)}]{sutherland_beautiful_2004}%
    \BibitemOpen
    \bibfield  {author} {\bibinfo {author} {\bibfnamefont {B.}~\bibnamefont
    {Sutherland}},\ }\href
    {https://www.worldscientific.com/worldscibooks/10.1142/5552#t=aboutBook}
    {\emph {\bibinfo {title} {Beautiful Models: 70 Years of Exactly Solved
    Quantum Many-Body Problems}}}\ (\bibinfo  {publisher} {World Scientific,
    Singapore},\ \bibinfo {year} {2004})\BibitemShut {NoStop}%
  \bibitem [{\citenamefont {Shiba}(1980)}]{shiba_quantization_1980}%
    \BibitemOpen
    \bibfield  {author} {\bibinfo {author} {\bibfnamefont {H.}~\bibnamefont
    {Shiba}},\ }\bibfield  {title} {\bibinfo {title} {Quantization of {{Magnetic
    Excitation Continuum Due}} to {{Interchain Coupling}} in {{Nearly
    One-Dimensional Ising-Like Antiferromagnets}}},\ }\href
    {https://doi.org/10.1143/PTP.64.466} {\bibfield  {journal} {\bibinfo
    {journal} {Prog. Theoretical Phys.}\ }\textbf {\bibinfo {volume} {64}},\
    \bibinfo {pages} {466} (\bibinfo {year} {1980})}\BibitemShut {NoStop}%
  \bibitem [{\citenamefont {Coldea}\ \emph {et~al.}(2010)\citenamefont {Coldea},
    \citenamefont {Tennant}, \citenamefont {Wheeler}, \citenamefont {Wawrzynska},
    \citenamefont {Prabhakaran}, \citenamefont {Telling}, \citenamefont
    {Habicht}, \citenamefont {Smeibidl},\ and\ \citenamefont
    {Kiefer}}]{coldea_quantum_2010}%
    \BibitemOpen
    \bibfield  {author} {\bibinfo {author} {\bibfnamefont {R.}~\bibnamefont
    {Coldea}}, \bibinfo {author} {\bibfnamefont {D.~A.}\ \bibnamefont {Tennant}},
    \bibinfo {author} {\bibfnamefont {E.~M.}\ \bibnamefont {Wheeler}}, \bibinfo
    {author} {\bibfnamefont {E.}~\bibnamefont {Wawrzynska}}, \bibinfo {author}
    {\bibfnamefont {D.}~\bibnamefont {Prabhakaran}}, \bibinfo {author}
    {\bibfnamefont {M.}~\bibnamefont {Telling}}, \bibinfo {author} {\bibfnamefont
    {K.}~\bibnamefont {Habicht}}, \bibinfo {author} {\bibfnamefont
    {P.}~\bibnamefont {Smeibidl}},\ and\ \bibinfo {author} {\bibfnamefont
    {K.}~\bibnamefont {Kiefer}},\ }\bibfield  {title} {\bibinfo {title} {Quantum
    {{Criticality}} in an {{Ising Chain}}: {{Experimental Evidence}} for
    {{Emergent E8 Symmetry}}},\ }\href {https://doi.org/10.1126/science.1180085}
    {\bibfield  {journal} {\bibinfo  {journal} {Science}\ }\textbf {\bibinfo
    {volume} {327}},\ \bibinfo {pages} {177} (\bibinfo {year}
    {2010})}\BibitemShut {NoStop}%
  \bibitem [{\citenamefont {Grenier}\ \emph {et~al.}(2015)\citenamefont
    {Grenier}, \citenamefont {Petit}, \citenamefont {Simonet}, \citenamefont
    {Can{\'e}vet}, \citenamefont {Regnault}, \citenamefont {Raymond},
    \citenamefont {Canals}, \citenamefont {Berthier},\ and\ \citenamefont
    {Lejay}}]{grenier_longitudinal_2015}%
    \BibitemOpen
    \bibfield  {author} {\bibinfo {author} {\bibfnamefont {B.}~\bibnamefont
    {Grenier}}, \bibinfo {author} {\bibfnamefont {S.}~\bibnamefont {Petit}},
    \bibinfo {author} {\bibfnamefont {V.}~\bibnamefont {Simonet}}, \bibinfo
    {author} {\bibfnamefont {E.}~\bibnamefont {Can{\'e}vet}}, \bibinfo {author}
    {\bibfnamefont {L.-P.}\ \bibnamefont {Regnault}}, \bibinfo {author}
    {\bibfnamefont {S.}~\bibnamefont {Raymond}}, \bibinfo {author} {\bibfnamefont
    {B.}~\bibnamefont {Canals}}, \bibinfo {author} {\bibfnamefont
    {C.}~\bibnamefont {Berthier}},\ and\ \bibinfo {author} {\bibfnamefont
    {P.}~\bibnamefont {Lejay}},\ }\bibfield  {title} {\bibinfo {title}
    {Longitudinal and {{Transverse Zeeman Ladders}} in the {{Ising-Like Chain
    Antiferromagnet}} {BaCo$_2$V$_2$O$_8$}},\ }\href
    {https://doi.org/10.1103/PhysRevLett.114.017201} {\bibfield  {journal}
    {\bibinfo  {journal} {Phys. Rev. Lett.}\ }\textbf {\bibinfo {volume} {114}},\
    \bibinfo {pages} {017201} (\bibinfo {year} {2015})}\BibitemShut {NoStop}%
  \bibitem [{\citenamefont {Mena}\ \emph {et~al.}(2020)\citenamefont {Mena},
    \citenamefont {H{\"a}nni}, \citenamefont {Ward}, \citenamefont
    {Hirtenlechner}, \citenamefont {Bewley}, \citenamefont {Hubig}, \citenamefont
    {Schollw{\"o}ck}, \citenamefont {Normand}, \citenamefont {Kr{\"a}mer},
    \citenamefont {McMorrow},\ and\ \citenamefont
    {R{\"u}egg}}]{mena_thermal_2020}%
    \BibitemOpen
    \bibfield  {author} {\bibinfo {author} {\bibfnamefont {M.}~\bibnamefont
    {Mena}}, \bibinfo {author} {\bibfnamefont {N.}~\bibnamefont {H{\"a}nni}},
    \bibinfo {author} {\bibfnamefont {S.}~\bibnamefont {Ward}}, \bibinfo {author}
    {\bibfnamefont {E.}~\bibnamefont {Hirtenlechner}}, \bibinfo {author}
    {\bibfnamefont {R.}~\bibnamefont {Bewley}}, \bibinfo {author} {\bibfnamefont
    {C.}~\bibnamefont {Hubig}}, \bibinfo {author} {\bibfnamefont
    {U.}~\bibnamefont {Schollw{\"o}ck}}, \bibinfo {author} {\bibfnamefont
    {B.}~\bibnamefont {Normand}}, \bibinfo {author} {\bibfnamefont {K.~W.}\
    \bibnamefont {Kr{\"a}mer}}, \bibinfo {author} {\bibfnamefont {D.~F.}\
    \bibnamefont {McMorrow}},\ and\ \bibinfo {author} {\bibfnamefont {{\relax
    Ch}.}~\bibnamefont {R{\"u}egg}},\ }\bibfield  {title} {\bibinfo {title}
    {Thermal {{Control}} of {{Spin Excitations}} in the {{Coupled Ising-Chain
    Material}} {RbCoCl$_3$}},\ }\href
    {https://doi.org/10.1103/PhysRevLett.124.257201} {\bibfield  {journal}
    {\bibinfo  {journal} {Phys. Rev. Lett.}\ }\textbf {\bibinfo {volume} {124}},\
    \bibinfo {pages} {257201} (\bibinfo {year} {2020})}\BibitemShut {NoStop}%
  \bibitem [{\citenamefont {Lane}\ \emph {et~al.}(2020)\citenamefont {Lane},
    \citenamefont {Stock}, \citenamefont {Cheong}, \citenamefont {Demmel},
    \citenamefont {Ewings},\ and\ \citenamefont
    {Kr{\"u}ger}}]{lane_nonlinear_2020}%
    \BibitemOpen
    \bibfield  {author} {\bibinfo {author} {\bibfnamefont {H.}~\bibnamefont
    {Lane}}, \bibinfo {author} {\bibfnamefont {C.}~\bibnamefont {Stock}},
    \bibinfo {author} {\bibfnamefont {S.-W.}\ \bibnamefont {Cheong}}, \bibinfo
    {author} {\bibfnamefont {F.}~\bibnamefont {Demmel}}, \bibinfo {author}
    {\bibfnamefont {R.~A.}\ \bibnamefont {Ewings}},\ and\ \bibinfo {author}
    {\bibfnamefont {F.}~\bibnamefont {Kr{\"u}ger}},\ }\bibfield  {title}
    {\bibinfo {title} {Nonlinear soliton confinement in weakly coupled
    antiferromagnetic spin chains},\ }\href
    {https://doi.org/10.1103/PhysRevB.102.024437} {\bibfield  {journal} {\bibinfo
     {journal} {Phys. Rev. B}\ }\textbf {\bibinfo {volume} {102}},\ \bibinfo
    {pages} {024437} (\bibinfo {year} {2020})}\BibitemShut {NoStop}%
  \bibitem [{\citenamefont {Karbach}\ and\ \citenamefont
    {M{\"u}ller}(2000)}]{karbach_line_2000}%
    \BibitemOpen
    \bibfield  {author} {\bibinfo {author} {\bibfnamefont {M.}~\bibnamefont
    {Karbach}}\ and\ \bibinfo {author} {\bibfnamefont {G.}~\bibnamefont
    {M{\"u}ller}},\ }\bibfield  {title} {\bibinfo {title} {Line-shape predictions
    via {{Bethe}} ansatz for the one-dimensional spin-{$\frac{1}{2}$}
    {{Heisenberg}} antiferromagnet in a magnetic field},\ }\href
    {https://doi.org/10.1103/PhysRevB.62.14871} {\bibfield  {journal} {\bibinfo
    {journal} {Phys. Rev. B}\ }\textbf {\bibinfo {volume} {62}},\ \bibinfo
    {pages} {14871} (\bibinfo {year} {2000})}\BibitemShut {NoStop}%
  \bibitem [{\citenamefont {Karbach}\ \emph {et~al.}(2002)\citenamefont
    {Karbach}, \citenamefont {Biegel},\ and\ \citenamefont
    {M{\"u}ller}}]{karbach_quasiparticles_2002}%
    \BibitemOpen
    \bibfield  {author} {\bibinfo {author} {\bibfnamefont {M.}~\bibnamefont
    {Karbach}}, \bibinfo {author} {\bibfnamefont {D.}~\bibnamefont {Biegel}},\
    and\ \bibinfo {author} {\bibfnamefont {G.}~\bibnamefont {M{\"u}ller}},\
    }\bibfield  {title} {\bibinfo {title} {Quasiparticles governing the
    zero-temperature dynamics of the one-dimensional spin-1/2 {{Heisenberg}}
    antiferromagnet in a magnetic field},\ }\href
    {https://doi.org/10.1103/PhysRevB.66.054405} {\bibfield  {journal} {\bibinfo
    {journal} {Phys. Rev. B}\ }\textbf {\bibinfo {volume} {66}},\ \bibinfo
    {pages} {054405} (\bibinfo {year} {2002})}\BibitemShut {NoStop}%
  \bibitem [{\citenamefont {Wang}\ \emph {et~al.}(2018)\citenamefont {Wang},
    \citenamefont {Wu}, \citenamefont {Yang}, \citenamefont {Bera}, \citenamefont
    {Kamenskyi}, \citenamefont {Islam}, \citenamefont {Xu}, \citenamefont {Law},
    \citenamefont {Lake}, \citenamefont {Wu},\ and\ \citenamefont
    {Loidl}}]{wang_experimental_2018}%
    \BibitemOpen
    \bibfield  {author} {\bibinfo {author} {\bibfnamefont {Z.}~\bibnamefont
    {Wang}}, \bibinfo {author} {\bibfnamefont {J.}~\bibnamefont {Wu}}, \bibinfo
    {author} {\bibfnamefont {W.}~\bibnamefont {Yang}}, \bibinfo {author}
    {\bibfnamefont {A.~K.}\ \bibnamefont {Bera}}, \bibinfo {author}
    {\bibfnamefont {D.}~\bibnamefont {Kamenskyi}}, \bibinfo {author}
    {\bibfnamefont {A.~T. M.~N.}\ \bibnamefont {Islam}}, \bibinfo {author}
    {\bibfnamefont {S.}~\bibnamefont {Xu}}, \bibinfo {author} {\bibfnamefont
    {J.~M.}\ \bibnamefont {Law}}, \bibinfo {author} {\bibfnamefont
    {B.}~\bibnamefont {Lake}}, \bibinfo {author} {\bibfnamefont {C.}~\bibnamefont
    {Wu}},\ and\ \bibinfo {author} {\bibfnamefont {A.}~\bibnamefont {Loidl}},\
    }\bibfield  {title} {\bibinfo {title} {Experimental observation of {{Bethe}}
    strings},\ }\href@noop {} {\bibfield  {journal} {\bibinfo  {journal}
    {Nature}\ }\textbf {\bibinfo {volume} {554}},\ \bibinfo {pages} {219}
    (\bibinfo {year} {2018})}\BibitemShut {NoStop}%
  \bibitem [{\citenamefont {Bera}\ \emph {et~al.}(2020)\citenamefont {Bera},
    \citenamefont {Wu}, \citenamefont {Yang}, \citenamefont {Bewley},
    \citenamefont {Boehm}, \citenamefont {Xu}, \citenamefont {Bartkowiak},
    \citenamefont {Prokhnenko}, \citenamefont {Klemke}, \citenamefont {Islam},
    \citenamefont {Law}, \citenamefont {Wang},\ and\ \citenamefont
    {Lake}}]{bera_dispersions_2020}%
    \BibitemOpen
    \bibfield  {author} {\bibinfo {author} {\bibfnamefont {A.~K.}\ \bibnamefont
    {Bera}}, \bibinfo {author} {\bibfnamefont {J.}~\bibnamefont {Wu}}, \bibinfo
    {author} {\bibfnamefont {W.}~\bibnamefont {Yang}}, \bibinfo {author}
    {\bibfnamefont {R.}~\bibnamefont {Bewley}}, \bibinfo {author} {\bibfnamefont
    {M.}~\bibnamefont {Boehm}}, \bibinfo {author} {\bibfnamefont
    {J.}~\bibnamefont {Xu}}, \bibinfo {author} {\bibfnamefont {M.}~\bibnamefont
    {Bartkowiak}}, \bibinfo {author} {\bibfnamefont {O.}~\bibnamefont
    {Prokhnenko}}, \bibinfo {author} {\bibfnamefont {B.}~\bibnamefont {Klemke}},
    \bibinfo {author} {\bibfnamefont {A.~T. M.~N.}\ \bibnamefont {Islam}},
    \bibinfo {author} {\bibfnamefont {J.~M.}\ \bibnamefont {Law}}, \bibinfo
    {author} {\bibfnamefont {Z.}~\bibnamefont {Wang}},\ and\ \bibinfo {author}
    {\bibfnamefont {B.}~\bibnamefont {Lake}},\ }\bibfield  {title} {\bibinfo
    {title} {Dispersions of many-body {{Bethe}} strings},\ }\href
    {https://doi.org/10.1038/s41567-020-0835-7} {\bibfield  {journal} {\bibinfo
    {journal} {Nat. Phys.}\ }\textbf {\bibinfo {volume} {16}},\ \bibinfo {pages}
    {625} (\bibinfo {year} {2020})}\BibitemShut {NoStop}%
  \bibitem [{\citenamefont {Kaul}\ \emph {et~al.}(2003)\citenamefont {Kaul},
    \citenamefont {Rosner}, \citenamefont {Yushankhai}, \citenamefont
    {Sichelschmidt}, \citenamefont {Shpanchenko},\ and\ \citenamefont
    {Geibel}}]{kaul_sr_2003}%
    \BibitemOpen
    \bibfield  {author} {\bibinfo {author} {\bibfnamefont {E.~E.}\ \bibnamefont
    {Kaul}}, \bibinfo {author} {\bibfnamefont {H.}~\bibnamefont {Rosner}},
    \bibinfo {author} {\bibfnamefont {V.}~\bibnamefont {Yushankhai}}, \bibinfo
    {author} {\bibfnamefont {J.}~\bibnamefont {Sichelschmidt}}, \bibinfo {author}
    {\bibfnamefont {R.~V.}\ \bibnamefont {Shpanchenko}},\ and\ \bibinfo {author}
    {\bibfnamefont {C.}~\bibnamefont {Geibel}},\ }\bibfield  {title} {\bibinfo
    {title} {{Sr$_2$V$_3$O$_9$} and {Ba$_2$V$_3$O$_9$}: {{Quasi-one-dimensional}}
    spin-systems with an anomalous low temperature susceptibility},\ }\href
    {https://doi.org/10.1103/PhysRevB.67.174417} {\bibfield  {journal} {\bibinfo
    {journal} {Phys. Rev. B}\ }\textbf {\bibinfo {volume} {67}},\ \bibinfo
    {pages} {174417} (\bibinfo {year} {2003})}\BibitemShut {NoStop}%
  \bibitem [{\citenamefont {Ivanshin}\ \emph {et~al.}(2003)\citenamefont
    {Ivanshin}, \citenamefont {Yushankhai}, \citenamefont {Sichelschmidt},
    \citenamefont {Zakharov}, \citenamefont {Kaul},\ and\ \citenamefont
    {Geibel}}]{ivanshin_esr_2003}%
    \BibitemOpen
    \bibfield  {author} {\bibinfo {author} {\bibfnamefont {V.~A.}\ \bibnamefont
    {Ivanshin}}, \bibinfo {author} {\bibfnamefont {V.}~\bibnamefont
    {Yushankhai}}, \bibinfo {author} {\bibfnamefont {J.}~\bibnamefont
    {Sichelschmidt}}, \bibinfo {author} {\bibfnamefont {D.~V.}\ \bibnamefont
    {Zakharov}}, \bibinfo {author} {\bibfnamefont {E.~E.}\ \bibnamefont {Kaul}},\
    and\ \bibinfo {author} {\bibfnamefont {C.}~\bibnamefont {Geibel}},\
    }\bibfield  {title} {\bibinfo {title} {{ESR} study of the anisotropic
    exchange in the quasi-one-dimensional antiferromagnet {Sr$_2$V$_3$O$_9$}},\
    }\href {https://doi.org/10.1103/PhysRevB.68.064404} {\bibfield  {journal}
    {\bibinfo  {journal} {Phys. Rev. B}\ }\textbf {\bibinfo {volume} {68}},\
    \bibinfo {pages} {064404} (\bibinfo {year} {2003})}\BibitemShut {NoStop}%
  \bibitem [{\citenamefont {Mentre}\ \emph {et~al.}(1998)\citenamefont {Mentre},
    \citenamefont {Dhaussy}, \citenamefont {Abraham},\ and\ \citenamefont
    {Steinfink}}]{mentre_structural_1998}%
    \BibitemOpen
    \bibfield  {author} {\bibinfo {author} {\bibfnamefont {O.}~\bibnamefont
    {Mentre}}, \bibinfo {author} {\bibfnamefont {A.-C.}\ \bibnamefont {Dhaussy}},
    \bibinfo {author} {\bibfnamefont {F.}~\bibnamefont {Abraham}},\ and\ \bibinfo
    {author} {\bibfnamefont {H.}~\bibnamefont {Steinfink}},\ }\bibfield  {title}
    {\bibinfo {title} {Structural, infrared, and magnetic characterization of the
    solid solution series {Sr$_{2-x}$Pb$_x$(VO)(VO$_4$)$_2$}; {Evidence} of the
    {Pb$^{2+}$} $6s^2$ lone pair stereochemical effect},\ }\href
    {https://doi.org/https://doi.org/10.1006/jssc.1998.7910} {\bibfield
    {journal} {\bibinfo  {journal} {J. Solid State Chem.}\ }\textbf {\bibinfo
    {volume} {140}},\ \bibinfo {pages} {417} (\bibinfo {year}
    {1998})}\BibitemShut {NoStop}%
  \bibitem [{\citenamefont {Kawamata}\ \emph {et~al.}(2014)\citenamefont
    {Kawamata}, \citenamefont {Uesaka}, \citenamefont {Sato}, \citenamefont
    {Naruse}, \citenamefont {Kudo}, \citenamefont {Kobayashi},\ and\
    \citenamefont {Koike}}]{kawamata_thermal_2014}%
    \BibitemOpen
    \bibfield  {author} {\bibinfo {author} {\bibfnamefont {T.}~\bibnamefont
    {Kawamata}}, \bibinfo {author} {\bibfnamefont {M.}~\bibnamefont {Uesaka}},
    \bibinfo {author} {\bibfnamefont {M.}~\bibnamefont {Sato}}, \bibinfo {author}
    {\bibfnamefont {K.}~\bibnamefont {Naruse}}, \bibinfo {author} {\bibfnamefont
    {K.}~\bibnamefont {Kudo}}, \bibinfo {author} {\bibfnamefont {N.}~\bibnamefont
    {Kobayashi}},\ and\ \bibinfo {author} {\bibfnamefont {Y.}~\bibnamefont
    {Koike}},\ }\bibfield  {title} {\bibinfo {title} {Thermal {{Conductivity}}
    due to {{Spinons}} in the {{One-Dimensional Quantum Spin System
    {Sr$_2$V$_3$O$_9$}}}},\ }\href {https://doi.org/10.7566/JPSJ.83.054601}
    {\bibfield  {journal} {\bibinfo  {journal} {J. Phys. Soc. Japan}\ }\textbf
    {\bibinfo {volume} {83}},\ \bibinfo {pages} {054601} (\bibinfo {year}
    {2014})}\BibitemShut {NoStop}%
  \bibitem [{\citenamefont {Rodr\'{\i}guez-Fortea}\ \emph
    {et~al.}(2010)\citenamefont {Rodr\'{\i}guez-Fortea}, \citenamefont {Llunell},
    \citenamefont {Alemany},\ and\ \citenamefont
    {Canadell}}]{rodriguez_first_2010}%
    \BibitemOpen
    \bibfield  {author} {\bibinfo {author} {\bibfnamefont {A.}~\bibnamefont
    {Rodr\'{\i}guez-Fortea}}, \bibinfo {author} {\bibfnamefont {M.}~\bibnamefont
    {Llunell}}, \bibinfo {author} {\bibfnamefont {P.}~\bibnamefont {Alemany}},\
    and\ \bibinfo {author} {\bibfnamefont {E.}~\bibnamefont {Canadell}},\
    }\bibfield  {title} {\bibinfo {title} {First-principles study of the
    interaction between paramagnetic {V$^{4+}$} centers through formally
    magnetically inactive {VO}$_{4}$ tetrahedra in the quasi-one-dimensional spin
    systems {Sr$_2$V$_3$O$_9$} and {Ba$_2$V$_3$O$_9$}},\ }\href
    {https://doi.org/10.1103/PhysRevB.82.134416} {\bibfield  {journal} {\bibinfo
    {journal} {Phys. Rev. B}\ }\textbf {\bibinfo {volume} {82}},\ \bibinfo
    {pages} {134416} (\bibinfo {year} {2010})}\BibitemShut {NoStop}%
  \bibitem [{\citenamefont {Uesaka}\ \emph {et~al.}(2010)\citenamefont {Uesaka},
    \citenamefont {Kawamata}, \citenamefont {Kaneko}, \citenamefont {Sato},
    \citenamefont {Kudo}, \citenamefont {Kobayashi},\ and\ \citenamefont
    {Koike}}]{uesaka_thermal_2010}%
    \BibitemOpen
    \bibfield  {author} {\bibinfo {author} {\bibfnamefont {M.}~\bibnamefont
    {Uesaka}}, \bibinfo {author} {\bibfnamefont {T.}~\bibnamefont {Kawamata}},
    \bibinfo {author} {\bibfnamefont {N.}~\bibnamefont {Kaneko}}, \bibinfo
    {author} {\bibfnamefont {M.}~\bibnamefont {Sato}}, \bibinfo {author}
    {\bibfnamefont {K.}~\bibnamefont {Kudo}}, \bibinfo {author} {\bibfnamefont
    {N.}~\bibnamefont {Kobayashi}},\ and\ \bibinfo {author} {\bibfnamefont
    {Y.}~\bibnamefont {Koike}},\ }\bibfield  {title} {\bibinfo {title} {Thermal
    conductivity of the quasi one-dimensional spin system {Sr$_2$V$_3$O$_9$}},\
    }\href {https://doi.org/10.1088/1742-6596/200/2/022068} {\bibfield  {journal}
    {\bibinfo  {journal} {J. Phys.: Conf. Series}\ }\textbf {\bibinfo {volume}
    {200}},\ \bibinfo {pages} {022068} (\bibinfo {year} {2010})}\BibitemShut
    {NoStop}%
  \bibitem [{\citenamefont {Arnold}\ \emph {et~al.}(2014)\citenamefont {Arnold},
    \citenamefont {Bilheux}, \citenamefont {Borreguero}, \citenamefont {Buts},
    \citenamefont {Campbell}, \citenamefont {Chapon}, \citenamefont {Doucet},
    \citenamefont {Draper}, \citenamefont {{Ferraz Leal}}, \citenamefont {Gigg},
    \citenamefont {Lynch}, \citenamefont {Markvardsen}, \citenamefont
    {Mikkelson}, \citenamefont {Mikkelson}, \citenamefont {Miller}, \citenamefont
    {Palmen}, \citenamefont {Parker}, \citenamefont {Passos}, \citenamefont
    {Perring}, \citenamefont {Peterson}, \citenamefont {Ren}, \citenamefont
    {Reuter}, \citenamefont {Savici}, \citenamefont {Taylor}, \citenamefont
    {Taylor}, \citenamefont {Tolchenov}, \citenamefont {Zhou},\ and\
    \citenamefont {Zikovsky}}]{arnold_mantid_2014}%
    \BibitemOpen
    \bibfield  {author} {\bibinfo {author} {\bibfnamefont {O.}~\bibnamefont
    {Arnold}}, \bibinfo {author} {\bibfnamefont {J.}~\bibnamefont {Bilheux}},
    \bibinfo {author} {\bibfnamefont {J.}~\bibnamefont {Borreguero}}, \bibinfo
    {author} {\bibfnamefont {A.}~\bibnamefont {Buts}}, \bibinfo {author}
    {\bibfnamefont {S.}~\bibnamefont {Campbell}}, \bibinfo {author}
    {\bibfnamefont {L.}~\bibnamefont {Chapon}}, \bibinfo {author} {\bibfnamefont
    {M.}~\bibnamefont {Doucet}}, \bibinfo {author} {\bibfnamefont
    {N.}~\bibnamefont {Draper}}, \bibinfo {author} {\bibfnamefont
    {R.}~\bibnamefont {{Ferraz Leal}}}, \bibinfo {author} {\bibfnamefont
    {M.}~\bibnamefont {Gigg}}, \bibinfo {author} {\bibfnamefont {V.}~\bibnamefont
    {Lynch}}, \bibinfo {author} {\bibfnamefont {A.}~\bibnamefont {Markvardsen}},
    \bibinfo {author} {\bibfnamefont {D.}~\bibnamefont {Mikkelson}}, \bibinfo
    {author} {\bibfnamefont {R.}~\bibnamefont {Mikkelson}}, \bibinfo {author}
    {\bibfnamefont {R.}~\bibnamefont {Miller}}, \bibinfo {author} {\bibfnamefont
    {K.}~\bibnamefont {Palmen}}, \bibinfo {author} {\bibfnamefont
    {P.}~\bibnamefont {Parker}}, \bibinfo {author} {\bibfnamefont
    {G.}~\bibnamefont {Passos}}, \bibinfo {author} {\bibfnamefont
    {T.}~\bibnamefont {Perring}}, \bibinfo {author} {\bibfnamefont
    {P.}~\bibnamefont {Peterson}}, \bibinfo {author} {\bibfnamefont
    {S.}~\bibnamefont {Ren}}, \bibinfo {author} {\bibfnamefont {M.}~\bibnamefont
    {Reuter}}, \bibinfo {author} {\bibfnamefont {A.}~\bibnamefont {Savici}},
    \bibinfo {author} {\bibfnamefont {J.}~\bibnamefont {Taylor}}, \bibinfo
    {author} {\bibfnamefont {R.}~\bibnamefont {Taylor}}, \bibinfo {author}
    {\bibfnamefont {R.}~\bibnamefont {Tolchenov}}, \bibinfo {author}
    {\bibfnamefont {W.}~\bibnamefont {Zhou}},\ and\ \bibinfo {author}
    {\bibfnamefont {J.}~\bibnamefont {Zikovsky}},\ }\bibfield  {title} {\bibinfo
    {title} {Mantid—{Data} analysis and visualization package for neutron
    scattering and $\mu${SR} experiments},\ }\href
    {https://doi.org/https://doi.org/10.1016/j.nima.2014.07.029} {\bibfield
    {journal} {\bibinfo  {journal} {Nucl. Instr. Methods Phys. Res. Sec. A}\
    }\textbf {\bibinfo {volume} {764}},\ \bibinfo {pages} {156 } (\bibinfo {year}
    {2014})}\BibitemShut {NoStop}%
  \bibitem [{\citenamefont {Caux}(2009)}]{caux_correlation_2009}%
    \BibitemOpen
    \bibfield  {author} {\bibinfo {author} {\bibfnamefont {J.-S.}\ \bibnamefont
    {Caux}},\ }\bibfield  {title} {\bibinfo {title} {{Correlation functions of
    integrable models: {A} description of the {ABACUS} algorithm}},\ }\href@noop
    {} {\bibfield  {journal} {\bibinfo  {journal} {J. Math. Phys.}\ }\textbf
    {\bibinfo {volume} {50}},\ \bibinfo {pages} {095214} (\bibinfo {year}
    {2009})}\BibitemShut {NoStop}%
  \bibitem [{\citenamefont {Caux}\ and\ \citenamefont
    {Hagemans}(2006)}]{caux_four_2006}%
    \BibitemOpen
    \bibfield  {author} {\bibinfo {author} {\bibfnamefont {J.-S.}\ \bibnamefont
    {Caux}}\ and\ \bibinfo {author} {\bibfnamefont {R.}~\bibnamefont
    {Hagemans}},\ }\bibfield  {title} {\bibinfo {title} {The four-spinon
    dynamical structure factor of the {Heisenberg} chain},\ }\href
    {https://doi.org/10.1088/1742-5468/2006/12/P12013} {\bibfield  {journal}
    {\bibinfo  {journal} {J. Stat. Mech.}\ ,\ \bibinfo {pages} {P12013}}
    (\bibinfo {year} {2006})}\BibitemShut {NoStop}%
  \bibitem [{\citenamefont {White}(1992)}]{white_density_1992}%
    \BibitemOpen
    \bibfield  {author} {\bibinfo {author} {\bibfnamefont {S.~R.}\ \bibnamefont
    {White}},\ }\bibfield  {title} {\bibinfo {title} {Density matrix formulation
    for quantum renormalization groups},\ }\href
    {https://doi.org/10.1103/PhysRevLett.69.2863} {\bibfield  {journal} {\bibinfo
     {journal} {Phys. Rev. Lett.}\ }\textbf {\bibinfo {volume} {69}},\ \bibinfo
    {pages} {2863} (\bibinfo {year} {1992})}\BibitemShut {NoStop}%
  \bibitem [{\citenamefont {White}(1993)}]{white_density_1993}%
    \BibitemOpen
    \bibfield  {author} {\bibinfo {author} {\bibfnamefont {S.~R.}\ \bibnamefont
    {White}},\ }\bibfield  {title} {\bibinfo {title} {Density-matrix algorithms
    for quantum renormalization groups},\ }\href
    {https://doi.org/10.1103/PhysRevB.48.10345} {\bibfield  {journal} {\bibinfo
    {journal} {Phys. Rev. B}\ }\textbf {\bibinfo {volume} {48}},\ \bibinfo
    {pages} {10345} (\bibinfo {year} {1993})}\BibitemShut {NoStop}%
  \bibitem [{\citenamefont {Alvarez}(2009)}]{alvarez_density_2009}%
    \BibitemOpen
    \bibfield  {author} {\bibinfo {author} {\bibfnamefont {G.}~\bibnamefont
    {Alvarez}},\ }\bibfield  {title} {\bibinfo {title} {The density matrix
    renormalization group for strongly correlated electron systems: A generic
    implementation},\ }\href
    {https://doi.org/https://doi.org/10.1016/j.cpc.2009.02.016} {\bibfield
    {journal} {\bibinfo  {journal} {Comput. Phys. Commun.}\ }\textbf {\bibinfo
    {volume} {180}},\ \bibinfo {pages} {1572} (\bibinfo {year}
    {2009})}\BibitemShut {NoStop}%
  \bibitem [{\citenamefont {K\"uhner}\ and\ \citenamefont
    {White}(1999)}]{kuhner_dynamical_1999}%
    \BibitemOpen
    \bibfield  {author} {\bibinfo {author} {\bibfnamefont {T.~D.}\ \bibnamefont
    {K\"uhner}}\ and\ \bibinfo {author} {\bibfnamefont {S.~R.}\ \bibnamefont
    {White}},\ }\bibfield  {title} {\bibinfo {title} {Dynamical correlation
    functions using the density matrix renormalization group},\ }\href
    {https://doi.org/10.1103/PhysRevB.60.335} {\bibfield  {journal} {\bibinfo
    {journal} {Phys. Rev. B}\ }\textbf {\bibinfo {volume} {60}},\ \bibinfo
    {pages} {335} (\bibinfo {year} {1999})}\BibitemShut {NoStop}%
  \bibitem [{\citenamefont {Nocera}\ and\ \citenamefont
    {Alvarez}(2016{\natexlab{a}})}]{nocera_spectral_2016}%
    \BibitemOpen
    \bibfield  {author} {\bibinfo {author} {\bibfnamefont {A.}~\bibnamefont
    {Nocera}}\ and\ \bibinfo {author} {\bibfnamefont {G.}~\bibnamefont
    {Alvarez}},\ }\bibfield  {title} {\bibinfo {title} {Spectral functions with
    the density matrix renormalization group: Krylov-space approach for
    correction vectors},\ }\href {https://doi.org/10.1103/PhysRevE.94.053308}
    {\bibfield  {journal} {\bibinfo  {journal} {Phys. Rev. E}\ }\textbf {\bibinfo
    {volume} {94}},\ \bibinfo {pages} {053308} (\bibinfo {year}
    {2016}{\natexlab{a}})}\BibitemShut {NoStop}%
  \bibitem [{\citenamefont {Feiguin}\ and\ \citenamefont
    {White}(2005)}]{feiguin_finite_2005}%
    \BibitemOpen
    \bibfield  {author} {\bibinfo {author} {\bibfnamefont {A.~E.}\ \bibnamefont
    {Feiguin}}\ and\ \bibinfo {author} {\bibfnamefont {S.~R.}\ \bibnamefont
    {White}},\ }\bibfield  {title} {\bibinfo {title} {Finite-temperature density
    matrix renormalization using an enlarged {Hilbert} space},\ }\href
    {https://doi.org/10.1103/PhysRevB.72.220401} {\bibfield  {journal} {\bibinfo
    {journal} {Phys. Rev. B}\ }\textbf {\bibinfo {volume} {72}},\ \bibinfo
    {pages} {220401(R)} (\bibinfo {year} {2005})}\BibitemShut {NoStop}%
  \bibitem [{\citenamefont {Feiguin}\ and\ \citenamefont
    {Fiete}(2010)}]{feiguin_spectral_2010}%
    \BibitemOpen
    \bibfield  {author} {\bibinfo {author} {\bibfnamefont {A.~E.}\ \bibnamefont
    {Feiguin}}\ and\ \bibinfo {author} {\bibfnamefont {G.~A.}\ \bibnamefont
    {Fiete}},\ }\bibfield  {title} {\bibinfo {title} {Spectral properties of a
    spin-incoherent {Luttinger} liquid},\ }\href
    {https://doi.org/10.1103/PhysRevB.81.075108} {\bibfield  {journal} {\bibinfo
    {journal} {Phys. Rev. B}\ }\textbf {\bibinfo {volume} {81}},\ \bibinfo
    {pages} {075108} (\bibinfo {year} {2010})}\BibitemShut {NoStop}%
  \bibitem [{\citenamefont {Nocera}\ and\ \citenamefont
    {Alvarez}(2016{\natexlab{b}})}]{nocera_symmetry_2016}%
    \BibitemOpen
    \bibfield  {author} {\bibinfo {author} {\bibfnamefont {A.}~\bibnamefont
    {Nocera}}\ and\ \bibinfo {author} {\bibfnamefont {G.}~\bibnamefont
    {Alvarez}},\ }\bibfield  {title} {\bibinfo {title} {Symmetry-conserving
    purification of quantum states within the density matrix renormalization
    group},\ }\href {https://doi.org/10.1103/PhysRevB.93.045137} {\bibfield
    {journal} {\bibinfo  {journal} {Phys. Rev. B}\ }\textbf {\bibinfo {volume}
    {93}},\ \bibinfo {pages} {045137} (\bibinfo {year}
    {2016}{\natexlab{b}})}\BibitemShut {NoStop}%
  \bibitem [{sup()}]{supp}%
    \BibitemOpen
    \href@noop {} {}\bibinfo {note} {See Supplemental Materials for discussions
    on the background scattering and interchain couplings, together with the
    input files and the calculation details of the DMRG
    calculations.}\BibitemShut {Stop}%
  \bibitem [{\citenamefont {Feyerherm}\ \emph {et~al.}(2000)\citenamefont
    {Feyerherm}, \citenamefont {Abens}, \citenamefont {Günther}, \citenamefont
    {Ishida}, \citenamefont {Meißner}, \citenamefont {Meschke}, \citenamefont
    {Nogami},\ and\ \citenamefont {Steiner}}]{Feyerherm_magnetic_2000}%
    \BibitemOpen
    \bibfield  {author} {\bibinfo {author} {\bibfnamefont {R.}~\bibnamefont
    {Feyerherm}}, \bibinfo {author} {\bibfnamefont {S.}~\bibnamefont {Abens}},
    \bibinfo {author} {\bibfnamefont {D.}~\bibnamefont {Günther}}, \bibinfo
    {author} {\bibfnamefont {T.}~\bibnamefont {Ishida}}, \bibinfo {author}
    {\bibfnamefont {M.}~\bibnamefont {Meißner}}, \bibinfo {author}
    {\bibfnamefont {M.}~\bibnamefont {Meschke}}, \bibinfo {author} {\bibfnamefont
    {T.}~\bibnamefont {Nogami}},\ and\ \bibinfo {author} {\bibfnamefont
    {M.}~\bibnamefont {Steiner}},\ }\bibfield  {title} {\bibinfo {title}
    {Magnetic-field induced gap and staggered susceptibility in the s = 1/2 chain
    [pm·cu(no3)2·(h2o)2]n (pm = pyrimidine)},\ }\href
    {https://doi.org/10.1088/0953-8984/12/39/312} {\bibfield  {journal} {\bibinfo
     {journal} {J. Phys.: Condensed Matt.}\ }\textbf {\bibinfo {volume} {12}},\
    \bibinfo {pages} {8495} (\bibinfo {year} {2000})}\BibitemShut {NoStop}%
  \bibitem [{\citenamefont {Dender}\ \emph {et~al.}(1997)\citenamefont {Dender},
    \citenamefont {Hammar}, \citenamefont {Reich}, \citenamefont {Broholm},\ and\
    \citenamefont {Aeppli}}]{dender_direct_1997}%
    \BibitemOpen
    \bibfield  {author} {\bibinfo {author} {\bibfnamefont {D.~C.}\ \bibnamefont
    {Dender}}, \bibinfo {author} {\bibfnamefont {P.~R.}\ \bibnamefont {Hammar}},
    \bibinfo {author} {\bibfnamefont {D.~H.}\ \bibnamefont {Reich}}, \bibinfo
    {author} {\bibfnamefont {C.}~\bibnamefont {Broholm}},\ and\ \bibinfo {author}
    {\bibfnamefont {G.}~\bibnamefont {Aeppli}},\ }\bibfield  {title} {\bibinfo
    {title} {Direct observation of field-induced incommensurate fluctuations in a
    one-dimensional ${S} = 1/2$ antiferromagnet},\ }\href
    {https://doi.org/10.1103/PhysRevLett.79.1750} {\bibfield  {journal} {\bibinfo
     {journal} {Phys. Rev. Lett.}\ }\textbf {\bibinfo {volume} {79}},\ \bibinfo
    {pages} {1750} (\bibinfo {year} {1997})}\BibitemShut {NoStop}%
  \bibitem [{\citenamefont {Oshikawa}\ and\ \citenamefont
    {Affleck}(1997)}]{oshikawa_field_1997}%
    \BibitemOpen
    \bibfield  {author} {\bibinfo {author} {\bibfnamefont {M.}~\bibnamefont
    {Oshikawa}}\ and\ \bibinfo {author} {\bibfnamefont {I.}~\bibnamefont
    {Affleck}},\ }\bibfield  {title} {\bibinfo {title} {Field-induced gap in
    ${S}= 1/2$ antiferromagnetic chains},\ }\href
    {https://doi.org/10.1103/PhysRevLett.79.2883} {\bibfield  {journal} {\bibinfo
     {journal} {Phys. Rev. Lett.}\ }\textbf {\bibinfo {volume} {79}},\ \bibinfo
    {pages} {2883} (\bibinfo {year} {1997})}\BibitemShut {NoStop}%
  \bibitem [{\citenamefont {M\"uller}\ \emph {et~al.}(1981)\citenamefont
    {M\"uller}, \citenamefont {Thomas}, \citenamefont {Beck},\ and\ \citenamefont
    {Bonner}}]{muller_quantum_1981}%
    \BibitemOpen
    \bibfield  {author} {\bibinfo {author} {\bibfnamefont {G.}~\bibnamefont
    {M\"uller}}, \bibinfo {author} {\bibfnamefont {H.}~\bibnamefont {Thomas}},
    \bibinfo {author} {\bibfnamefont {H.}~\bibnamefont {Beck}},\ and\ \bibinfo
    {author} {\bibfnamefont {J.~C.}\ \bibnamefont {Bonner}},\ }\bibfield  {title}
    {\bibinfo {title} {Quantum spin dynamics of the antiferromagnetic linear
    chain in zero and nonzero magnetic field},\ }\href
    {https://doi.org/10.1103/PhysRevB.24.1429} {\bibfield  {journal} {\bibinfo
    {journal} {Phys. Rev. B}\ }\textbf {\bibinfo {volume} {24}},\ \bibinfo
    {pages} {1429} (\bibinfo {year} {1981})}\BibitemShut {NoStop}%
  \bibitem [{\citenamefont {Tennant}\ \emph {et~al.}(1995)\citenamefont
    {Tennant}, \citenamefont {Cowley}, \citenamefont {Nagler},\ and\
    \citenamefont {Tsvelik}}]{tennant_measurement_1995}%
    \BibitemOpen
    \bibfield  {author} {\bibinfo {author} {\bibfnamefont {D.~A.}\ \bibnamefont
    {Tennant}}, \bibinfo {author} {\bibfnamefont {R.~A.}\ \bibnamefont {Cowley}},
    \bibinfo {author} {\bibfnamefont {S.~E.}\ \bibnamefont {Nagler}},\ and\
    \bibinfo {author} {\bibfnamefont {A.~M.}\ \bibnamefont {Tsvelik}},\
    }\bibfield  {title} {\bibinfo {title} {Measurement of the spin-excitation
    continuum in one-dimensional {KCuF$_3$} using neutron scattering},\
    }\href@noop {} {\bibfield  {journal} {\bibinfo  {journal} {Phys. Rev. B}\
    }\textbf {\bibinfo {volume} {52}},\ \bibinfo {pages} {13368} (\bibinfo {year}
    {1995})}\BibitemShut {NoStop}%
  \bibitem [{\citenamefont {Lake}\ \emph {et~al.}(2005)\citenamefont {Lake},
    \citenamefont {Tennant}, \citenamefont {Frost},\ and\ \citenamefont
    {Nagler}}]{lake_quantum_2005}%
    \BibitemOpen
    \bibfield  {author} {\bibinfo {author} {\bibfnamefont {B.}~\bibnamefont
    {Lake}}, \bibinfo {author} {\bibfnamefont {D.~A.}\ \bibnamefont {Tennant}},
    \bibinfo {author} {\bibfnamefont {C.~D.}\ \bibnamefont {Frost}},\ and\
    \bibinfo {author} {\bibfnamefont {S.~E.}\ \bibnamefont {Nagler}},\ }\bibfield
     {title} {\bibinfo {title} {Quantum criticality and universal scaling of a
    quantum antiferromagnet},\ }\href@noop {} {\bibfield  {journal} {\bibinfo
    {journal} {Nat. Mater.}\ }\textbf {\bibinfo {volume} {4}},\ \bibinfo {pages}
    {329} (\bibinfo {year} {2005})}\BibitemShut {NoStop}%
  \bibitem [{\citenamefont {Mourigal}\ \emph {et~al.}(2013)\citenamefont
    {Mourigal}, \citenamefont {Enderle}, \citenamefont {Klopperpieper},
    \citenamefont {Caux}, \citenamefont {Stunault},\ and\ \citenamefont
    {Ronnow}}]{mourigal_fractional_2013}%
    \BibitemOpen
    \bibfield  {author} {\bibinfo {author} {\bibfnamefont {M.}~\bibnamefont
    {Mourigal}}, \bibinfo {author} {\bibfnamefont {M.}~\bibnamefont {Enderle}},
    \bibinfo {author} {\bibfnamefont {A.}~\bibnamefont {Klopperpieper}}, \bibinfo
    {author} {\bibfnamefont {J.-S.}\ \bibnamefont {Caux}}, \bibinfo {author}
    {\bibfnamefont {A.}~\bibnamefont {Stunault}},\ and\ \bibinfo {author}
    {\bibfnamefont {H.~M.}\ \bibnamefont {Ronnow}},\ }\bibfield  {title}
    {\bibinfo {title} {Fractional spinon excitations in the quantum
    {{Heisenberg}} antiferromagnetic chain},\ }\href
    {https://doi.org/10.1038/nphys2652} {\bibfield  {journal} {\bibinfo
    {journal} {Nat. Phys.}\ }\textbf {\bibinfo {volume} {9}},\ \bibinfo {pages}
    {435} (\bibinfo {year} {2013})}\BibitemShut {NoStop}%
  \bibitem [{\citenamefont {Wu}\ \emph {et~al.}(2019)\citenamefont {Wu},
    \citenamefont {Nikitin}, \citenamefont {Wang}, \citenamefont {Zhu},
    \citenamefont {Batista}, \citenamefont {Tsvelik}, \citenamefont {Samarakoon},
    \citenamefont {Tennant}, \citenamefont {Brando}, \citenamefont {Vasylechko},
    \citenamefont {Frontzek}, \citenamefont {Savici}, \citenamefont {Sala},
    \citenamefont {Ehlers}, \citenamefont {Christianson}, \citenamefont
    {Lumsden},\ and\ \citenamefont {Podlesnyak}}]{wu_tomonaga_2019}%
    \BibitemOpen
    \bibfield  {author} {\bibinfo {author} {\bibfnamefont {L.~S.}\ \bibnamefont
    {Wu}}, \bibinfo {author} {\bibfnamefont {S.~E.}\ \bibnamefont {Nikitin}},
    \bibinfo {author} {\bibfnamefont {Z.}~\bibnamefont {Wang}}, \bibinfo {author}
    {\bibfnamefont {W.}~\bibnamefont {Zhu}}, \bibinfo {author} {\bibfnamefont
    {C.~D.}\ \bibnamefont {Batista}}, \bibinfo {author} {\bibfnamefont {A.~M.}\
    \bibnamefont {Tsvelik}}, \bibinfo {author} {\bibfnamefont {A.~M.}\
    \bibnamefont {Samarakoon}}, \bibinfo {author} {\bibfnamefont {D.~A.}\
    \bibnamefont {Tennant}}, \bibinfo {author} {\bibfnamefont {M.}~\bibnamefont
    {Brando}}, \bibinfo {author} {\bibfnamefont {L.}~\bibnamefont {Vasylechko}},
    \bibinfo {author} {\bibfnamefont {M.}~\bibnamefont {Frontzek}}, \bibinfo
    {author} {\bibfnamefont {A.~T.}\ \bibnamefont {Savici}}, \bibinfo {author}
    {\bibfnamefont {G.}~\bibnamefont {Sala}}, \bibinfo {author} {\bibfnamefont
    {G.}~\bibnamefont {Ehlers}}, \bibinfo {author} {\bibfnamefont {A.~D.}\
    \bibnamefont {Christianson}}, \bibinfo {author} {\bibfnamefont {M.~D.}\
    \bibnamefont {Lumsden}},\ and\ \bibinfo {author} {\bibfnamefont
    {A.}~\bibnamefont {Podlesnyak}},\ }\bibfield  {title} {\bibinfo {title}
    {Tomonaga-{Luttinger} liquid behavior and spinon confinement in
    {YbAlO$_3$}},\ }\href {https://doi.org/10.1038/s41467-019-08485-7} {\bibfield
     {journal} {\bibinfo  {journal} {Nat. Commun.}\ }\textbf {\bibinfo {volume}
    {10}},\ \bibinfo {pages} {698} (\bibinfo {year} {2019})}\BibitemShut
    {NoStop}%
  \bibitem [{\citenamefont {Bougourzi}\ \emph {et~al.}(1996)\citenamefont
    {Bougourzi}, \citenamefont {Couture},\ and\ \citenamefont
    {Kacir}}]{bougourzi_exact_1996}%
    \BibitemOpen
    \bibfield  {author} {\bibinfo {author} {\bibfnamefont {A.~H.}\ \bibnamefont
    {Bougourzi}}, \bibinfo {author} {\bibfnamefont {M.}~\bibnamefont {Couture}},\
    and\ \bibinfo {author} {\bibfnamefont {M.}~\bibnamefont {Kacir}},\ }\bibfield
     {title} {\bibinfo {title} {Exact two-spinon dynamical correlation function
    of the one-dimensional {Heisenberg} model},\ }\href
    {https://doi.org/10.1103/PhysRevB.54.R12669} {\bibfield  {journal} {\bibinfo
    {journal} {Phys. Rev. B}\ }\textbf {\bibinfo {volume} {54}},\ \bibinfo
    {pages} {R12669} (\bibinfo {year} {1996})}\BibitemShut {NoStop}%
  \bibitem [{\citenamefont {Karbach}\ \emph {et~al.}(1997)\citenamefont
    {Karbach}, \citenamefont {M\"uller}, \citenamefont {Bougourzi}, \citenamefont
    {Fledderjohann},\ and\ \citenamefont {M\"utter}}]{karbach_two_1997}%
    \BibitemOpen
    \bibfield  {author} {\bibinfo {author} {\bibfnamefont {M.}~\bibnamefont
    {Karbach}}, \bibinfo {author} {\bibfnamefont {G.}~\bibnamefont {M\"uller}},
    \bibinfo {author} {\bibfnamefont {A.~H.}\ \bibnamefont {Bougourzi}}, \bibinfo
    {author} {\bibfnamefont {A.}~\bibnamefont {Fledderjohann}},\ and\ \bibinfo
    {author} {\bibfnamefont {K.-H.}\ \bibnamefont {M\"utter}},\ }\bibfield
    {title} {\bibinfo {title} {Two-spinon dynamic structure factor of the
    one-dimensional ${S}=\frac{1}{2}$ {Heisenberg} antiferromagnet},\ }\href
    {https://doi.org/10.1103/PhysRevB.55.12510} {\bibfield  {journal} {\bibinfo
    {journal} {Phys. Rev. B}\ }\textbf {\bibinfo {volume} {55}},\ \bibinfo
    {pages} {12510} (\bibinfo {year} {1997})}\BibitemShut {NoStop}%
  \bibitem [{\citenamefont {Lake}\ \emph {et~al.}(2013)\citenamefont {Lake},
    \citenamefont {Tennant}, \citenamefont {Caux}, \citenamefont {Barthel},
    \citenamefont {Schollw{\"o}ck}, \citenamefont {Nagler},\ and\ \citenamefont
    {Frost}}]{lake_multispinon_2013}%
    \BibitemOpen
    \bibfield  {author} {\bibinfo {author} {\bibfnamefont {B.}~\bibnamefont
    {Lake}}, \bibinfo {author} {\bibfnamefont {D.~A.}\ \bibnamefont {Tennant}},
    \bibinfo {author} {\bibfnamefont {J.-S.}\ \bibnamefont {Caux}}, \bibinfo
    {author} {\bibfnamefont {T.}~\bibnamefont {Barthel}}, \bibinfo {author}
    {\bibfnamefont {U.}~\bibnamefont {Schollw{\"o}ck}}, \bibinfo {author}
    {\bibfnamefont {S.~E.}\ \bibnamefont {Nagler}},\ and\ \bibinfo {author}
    {\bibfnamefont {C.~D.}\ \bibnamefont {Frost}},\ }\bibfield  {title} {\bibinfo
    {title} {Multispinon {{Continua}} at {{Zero}} and {{Finite Temperature}} in a
    {{Near-Ideal Heisenberg Chain}}},\ }\href
    {https://doi.org/10.1103/PhysRevLett.111.137205} {\bibfield  {journal}
    {\bibinfo  {journal} {Phys. Rev. Lett.}\ }\textbf {\bibinfo {volume} {111}},\
    \bibinfo {pages} {137205} (\bibinfo {year} {2013})}\BibitemShut {NoStop}%
  \bibitem [{\citenamefont {Schulz}(1996)}]{schulz_dynamics_1996}%
    \BibitemOpen
    \bibfield  {author} {\bibinfo {author} {\bibfnamefont {H.~J.}\ \bibnamefont
    {Schulz}},\ }\bibfield  {title} {\bibinfo {title} {Dynamics of coupled
    quantum spin chains},\ }\href {https://doi.org/10.1103/PhysRevLett.77.2790}
    {\bibfield  {journal} {\bibinfo  {journal} {Phys. Rev. Lett.}\ }\textbf
    {\bibinfo {volume} {77}},\ \bibinfo {pages} {2790} (\bibinfo {year}
    {1996})}\BibitemShut {NoStop}%
  \bibitem [{\citenamefont {Starykh}\ \emph {et~al.}(1997)\citenamefont
    {Starykh}, \citenamefont {Sandvik},\ and\ \citenamefont
    {Singh}}]{starykh_dynamics_1997}%
    \BibitemOpen
    \bibfield  {author} {\bibinfo {author} {\bibfnamefont {O.~A.}\ \bibnamefont
    {Starykh}}, \bibinfo {author} {\bibfnamefont {A.~W.}\ \bibnamefont
    {Sandvik}},\ and\ \bibinfo {author} {\bibfnamefont {R.~R.~P.}\ \bibnamefont
    {Singh}},\ }\bibfield  {title} {\bibinfo {title} {Dynamics of the
    spin-{$\frac{1}{2}$} {Heisenberg} chain at intermediate temperatures},\
    }\href {https://doi.org/10.1103/PhysRevB.55.14953} {\bibfield  {journal}
    {\bibinfo  {journal} {Phys. Rev. B}\ }\textbf {\bibinfo {volume} {55}},\
    \bibinfo {pages} {14953} (\bibinfo {year} {1997})}\BibitemShut {NoStop}%
  \bibitem [{\citenamefont {Schulz}(1986)}]{schulz_phase_1986}%
    \BibitemOpen
    \bibfield  {author} {\bibinfo {author} {\bibfnamefont {H.~J.}\ \bibnamefont
    {Schulz}},\ }\bibfield  {title} {\bibinfo {title} {Phase diagrams and
    correlation exponents for quantum spin chains of arbitrary spin quantum
    number},\ }\href {https://doi.org/10.1103/PhysRevB.34.6372} {\bibfield
    {journal} {\bibinfo  {journal} {Phys. Rev. B}\ }\textbf {\bibinfo {volume}
    {34}},\ \bibinfo {pages} {6372} (\bibinfo {year} {1986})}\BibitemShut
    {NoStop}%
  \bibitem [{\citenamefont {Tennant}\ \emph {et~al.}(1993)\citenamefont
    {Tennant}, \citenamefont {Perring}, \citenamefont {Cowley},\ and\
    \citenamefont {Nagler}}]{tennant_unbound_1993}%
    \BibitemOpen
    \bibfield  {author} {\bibinfo {author} {\bibfnamefont {D.~A.}\ \bibnamefont
    {Tennant}}, \bibinfo {author} {\bibfnamefont {T.~G.}\ \bibnamefont
    {Perring}}, \bibinfo {author} {\bibfnamefont {R.~A.}\ \bibnamefont
    {Cowley}},\ and\ \bibinfo {author} {\bibfnamefont {S.~E.}\ \bibnamefont
    {Nagler}},\ }\bibfield  {title} {\bibinfo {title} {Unbound spinons in the
    {S}=1/2 antiferromagnetic chain {KCuF}$_3$},\ }\href
    {https://doi.org/10.1103/PhysRevLett.70.4003} {\bibfield  {journal} {\bibinfo
     {journal} {Phys. Rev. Lett.}\ }\textbf {\bibinfo {volume} {70}},\ \bibinfo
    {pages} {4003} (\bibinfo {year} {1993})}\BibitemShut {NoStop}%
  \end{thebibliography}

\begin{thebibliography}{99}
\bibitem{Scheie2021Witnessing} A. Scheie, Pontus Laurell, A. M. Samarakoon, B. Lake, S. E. Nagler, G. E. Granroth, S. Okamoto, G. Alvarez, and D. A. Tennant, {Witnessing entanglement in quantum magnets using neutron scattering,} \href {https://doi.org/10.1103/PhysRevB.103.224434}  {Phys. Rev. B \textbf {103}, 224434 (2021)}.
\bibitem{alvarez2009density} G. Alvarez, {The density matrix renormalization group for strongly correlated electron systems: A generic implementation,} \href {https://doi.org/10.1016/j.cpc.2009.02.016} {Comput. Phys. Commun. \textbf {180}, 1572 (2009)}.
\bibitem{Feiguin2005finite} A. E. Feiguin and S. R. White, {Finite-temperature density matrix renormalization using an enlarged Hilbert space,} \href {https://doi.org/10.1103/PhysRevB.72.220401} {Phys. Rev. B \textbf {72}, 220401 (2005)}.
\bibitem{Feiguin2010Spectral} A. E. Feiguin and G. A. Fiete, {Spectral properties of a spin-incoherent Luttinger liquid,} \href {https://doi.org/10.1103/PhysRevB.81.075108} {Phys. Rev. B \textbf {81}, 075108 (2010)}.
\bibitem{Nocera2016symmetry} A. Nocera and G. Alvarez, {Symmetry-conserving purification of quantum states within the density matrix renormalization group,} \href {https://doi.org/10.1103/PhysRevB.93.045137} {Phys. Rev. B \textbf {93}, 045137 (2016)}.

\end{thebibliography}
\end{document}